\documentclass{ifacconf}
\usepackage{amsmath}

\let\theoremstyle\relax   

\usepackage[utf8]{inputenc} 
\usepackage[T1]{fontenc}    
\usepackage{url}            
\usepackage{booktabs}   
\usepackage{amsmath}
\usepackage{amsfonts}
\usepackage{xcolor}
\usepackage{soul}
\usepackage{nicefrac}       
\usepackage{microtype}      
\usepackage{lipsum}
\usepackage{graphicx}       
\graphicspath{{media/}}     
 \usepackage{subcaption}
\usepackage[table]{xcolor}
\usepackage{stmaryrd}

\usepackage{algorithm} 
\usepackage{algpseudocode} 

\usepackage[utf8]{inputenc} 
\usepackage[T1]{fontenc}    
\usepackage{url}            
  
\usepackage{amsthm}
\usepackage{amsfonts}
\usepackage{graphicx}
\usepackage{xcolor}
\usepackage{epstopdf}
\usepackage{amssymb}
\usepackage{siunitx}
\usepackage{soul}
\usepackage{nicefrac}       
\usepackage{microtype}      
\usepackage{lipsum}
\usepackage{fancyhdr}  
\usepackage{booktabs} 

\usepackage{graphicx}       
\graphicspath{{media/}}     
\usepackage{stmaryrd}

\usepackage{algorithm} 
\usepackage{algpseudocode}

\newtheoremstyle{thmv2}
{}
{.1em}
{\itshape}
{}
{\bfseries}
{.}
{.1em}
{}
\theoremstyle{thmv2}

\newenvironment{tightdisplay}{%
  \begingroup
  \setlength\abovedisplayskip{6pt}%
  \setlength\belowdisplayskip{6pt}%
  \setlength\abovedisplayshortskip{4pt}%
  \setlength\belowdisplayshortskip{4pt}%
}{\endgroup}

\newtheorem{remark}{Remark}
\usepackage{mathtools}
\usepackage{tikz}
\usepackage{tikz}
\usetikzlibrary{positioning,arrows.meta}
\usetikzlibrary{shapes,arrows,automata}
\usetikzlibrary{intersections}
\usetikzlibrary{trees,positioning,arrows.meta,calc,3d,fit,backgrounds,matrix}

\usepackage{cancel}
\usepackage[mathscr]{eucal}
\usepackage{algpseudocode}
\usepackage{pgfplots}

\pgfplotscreatecolormap{BlueYellow}{%
	rgb255=(68,  1, 84)   
	rgb255=(59, 82,139)   
	rgb255=(33,145,140)   
	rgb255=(94,201, 98)   
	rgb255=(253,231,37)   
}

\newcommand{\union}{\cup}
\newcommand{\subphiSet}{\ensuremath{\scalebox{1.4}{$\alpha$}}}
\newcommand{\celldots}{\makebox[0.78cm][c]{$\ldots$}}

\newcommand{\spaceR}{\mathbb{R}} 

\newcommand{\spaceX}{\mathbb{X}} 
\newcommand{\spaceS}{\mathbb{S}}
\newcommand{\spaceU}{\mathbb{U}}

\newcommand{\spaceY}{\mathbb{Y}} 

\newcommand{\mbP}{\mathbf{P}}

\newcommand{\mdpM}{\mathbf{M}}
\newcommand{\Tr}{{\mathbb T}}

\newcommand{\s}[1]{s_{#1}} 
\newcommand{\AP}{\mathrm{AP}}

\newcommand{\word}{\boldsymbol{l}}
\newcommand{\M}{\mathbf{M}}

\newcommand{\DFA}{\mathcal A}
\newcommand{\DFAs}{{\DFA_{\phi}}}
\newcommand{\setnodes}{\mathrm{N}}

\newcommand{\trunc}{\mathbf{L}}
\newcommand\norm[1]{\left\lVert#1\right\rVert}

\newcommand{\subphi}{\alpha}

\newcommand{\valuemapping}{v}

\newcommand{\tensorV}{\mathscr{V}}
\newcommand{\qmapping}{\mathcal{L}_Q}

\newcommand{\optree}{\mathcal{T}}
\newcommand{\expectation}{\mathbb{E}}
\newcommand{\indicator}{\boldsymbol{1}}

\newcommand{\outerproduct}{\otimes}
\newcommand{\op}{\mathbf{T}}

\newcommand{\ruohan}[1]{{\color{blue}#1}}

\newtheorem{definition}{Definition}

\newtheorem{lemma}{Lemma}
\newtheorem{theorem}{Theorem}

\newtheorem{proposition}{Proposition}

\newcommand{\relation}{\mathscr{R}}

\renewcommand{\ni}{{n_i}}
\newcommand{\mi}{{m_i}}

\renewcommand{\ruohan}[1]{{\color{black}#1}}
\usepackage{subfig}
\usepackage{natbib}        
\usepackage{color}        
\makeatletter
\let\origthebibliography\thebibliography
\let\endorigthebibliography\endthebibliography
\renewcommand{\thebibliography}[1]{%
  \origthebibliography{#1}%
  \fontsize{9pt}{8.5pt}\selectfont
  \setlength{\labelwidth}{0pt}%
  \setlength{\labelsep}{0pt}%
}
\renewcommand{\endthebibliography}{\endorigthebibliography}
\renewcommand{\cite}{\citep}
\makeatother

\begin{document}
\begin{frontmatter}

\title{Correct-by-Design Control Synthesis of Stochastic Multi-agent Systems: a Robust Tensor-based Solution\thanksref{footnoteinfo}} 

\thanks[footnoteinfo]{This work is supported by the European projects SymAware under the grant number 101070802 and COVER under the grant number 101086228. This work is also supported by National Natural Science Foundation of China under the grant number 62533017. Corresponding author: \textit{Zhiyong Sun}.}

\author[First]{Ruohan Wang} 
\author[First]{Siyuan Liu} 
\author[Second]{Zhiyong Sun}
\author[First]{Sofie Haesaert}

\address[First]{Department of Electrical Engineering, TU Eindhoven, the Netherlands. Emails: $\{$r.wang2, s.liu5@tue.nl, s.haesaert$\}$@tue.nl.}
\address[Second]{Department of Mechanics and Engineering Science, Peking University, Beijing, China. Email: zhiyong.sun@pku.edu.cn.}

\begin{abstract}
	Discrete-time stochastic systems with continuous spaces are hard to verify and control, even with MDP abstractions, due to the curse of dimensionality. We propose an abstraction-based framework with robust dynamic programming mappings that deliver control strategies with provable lower bounds on temporal-logic satisfaction, quantified via approximate stochastic simulation relations. Exploiting decoupled dynamics, we reveal a Canonical Polyadic Decomposition tensor structure in value functions that makes dynamic programming scalable. The proposed method provides correct-by-design probabilistic guarantees for temporal logic specifications. We validate our results on continuous-state linear stochastic systems.
\end{abstract}
\end{frontmatter}

\section{Introduction}
 Advances in computational power have enabled the development of large-scale systems in safety-critical domains like smart grids and traffic management. These systems involve numerous agents with complex or uncertain dynamics. \ruohan{A common approach in the multi-agent control literature is to consider decoupled agent dynamics, with the multi-agent nature arising through shared objectives and coordinated synthesis \cite{wang2025distributed,wiltz2025parallelized}. At the same time, coordinating multiple agents increases system complexity and risk. Temporal-logic specifications provide a formal way to express safety objectives for stochastic multi-agent systems \cite{vlahakis2024probabilistic} modeled as general Markov decision processes (gMDP), in which stochasticity is governed by a Markov kernel and observed through output mappings \cite{schon2023verifying,cai2025safety}.}
In this framework, safety is encoded as a reachability problem on the product of the system and an automaton, which enables quantitative guarantees \cite{baier2008principles,belta2017formal}.
 To formalize and certify such guarantees, the continuous nature of the state (and possibly action) spaces motivates finite abstractions, with stochastic simulation relations providing certified bounds that link abstract and original models \cite{tabuada2009verification,desharnais2004metrics}.
 \par
 Correct-by-design control synthesis for stochastic systems with temporal-logic specifications faces the curse of dimensionality, as computational cost grows exponentially with system size, rendering synthesis intractable \cite{liu2021symbolic}. Prior efforts mitigate this mainly via scalable abstractions rather than accelerating the synthesis step itself \cite{haesaert2020robust,liu2021symbolic,schon2023verifying}. When compared to abstraction-free methods \citep{jagtap2020formal,vlahakis2024probabilistic}, abstraction-based approaches can surpass abstraction-free methods in quantitative accuracy while retaining certified guarantees \cite{lavaei2021compositional}. Yet, due to scalability bottlenecks in compositional analysis, such abstraction-driven techniques have not been realized at scale for multi-agent systems.


\par
Given a finite abstraction, the synthesis problem for a multi-agent system reduces to solving dynamic programming on a very large gMDP, with the computational bottleneck shifted from model construction to control synthesis. This shift motivates structure-exploiting accelerations of dynamic programming: by leveraging low-rank tensor approximations in dynamic programming problems \cite{rozada2024tensor}, the optimal solution can be approximated while reducing the computational complexity and mitigating scalability issues. For finite Markov Decision Processes (MDP), recent work \cite{wang2025unraveling} shows that a tree structure can be used to manage and prune the tensor rank in a provable fashion. 
\par 
\noindent {\bfseries Contributions.}\hspace{1mm} \ruohan{
This paper leverages low-rank tensors and establishes a first abstraction-based approach to scalable correct-by-design control synthesis for continuous-state multi-agent systems, thereby extending preliminary results in \cite{wang2025unraveling} which only held for finite-state systems. To this end we introduce two novel approaches to compute lower bounds on the satisfaction probability that are robust to the incurred abstraction errors.
Additionally, we demonstrate the effectiveness of the approach on multiple benchmarks.

}



\section{Framework and approach}
\subsection{\ruohan{Notations}}\vspace{-0.2cm}
Denote a Borel measurable space as $(\spaceS,\mathcal{B}(\spaceS))$ where $\spaceS$ is a set and $\mathcal{B}(\spaceS)$ is the Borel sigma-algebra on $\spaceS$. A probability measure $\mathbb{P}$ over $(\spaceS,\mathcal{B}(\spaceS))$ defines the probability space $(\spaceS,\mathcal{B}(\spaceS),\mathbb{P})$ and has realization $x\sim \mathbb{P}$. \ruohan{We denote the set of all probability measures for a given measurable space $(\spaceS,\mathcal{B}(\spaceS))$ as $\mathcal{P}(\spaceS)$.} We assume all such spaces $\spaceS$ are Polish. Let $d_{\mathbb{\spaceS}}:\spaceS \times\spaceS \to[0,\infty)$ be a \textit{metric} \footnote{We recall the axioms of a \textit{metric}: $d_\spaceS(x,y)\ge 0$, $d_\spaceS(x,y)=0 \Leftrightarrow x=y$, $d_\spaceS(x,y)=d_\spaceS(y,x)$, and $d_\spaceS(x,z)\le d_\spaceS(x,y)+d_\spaceS(y,z)$ for all $x,y,z\in\spaceS$.} on $\spaceS$. For vectors $x_i\in \mathbb{R}^{n_i}$ we define $\operatorname{col}(x_1,\ldots,x_N):=[x_1^\top\ldots x_N^\top]^\top \in\mathbb{R}^{\sum_{i=1}^N n_i}$. \ruohan{For a stochastic process with state variable $x$, we denote the expectation over the next state $x'$ conditioned on the current state $x$ as $\expectation_{x}$.
}

\vskip-.2cm
\subsection{System models: Stochastic difference equations}\vspace{-0.2cm}
In this paper, we consider a multi-agent system $\mathbf{M}$ composed of $N$ agents. Each agent  $\mathbf{M}^i$, $i\in\{1,\ldots,N\}$, is modeled with a
stochastic difference equation:
\begin{align}\label{eq:decoupled_hd_mdp}
		\left\{\begin{array}{ll}	x_i^+
		  &=    
			f_i(x_i,u_i) + 
			w_i \\
			y_i
		 &=  
			\ruohan{g_i}(x_i)  \end{array}\right. \forall i\in \{1,\ldots N\}
\end{align}
with state $x_i\in \spaceX_i\subset\spaceR^{\ni}$, the control input $u_i\in \spaceU^{i}\subset\spaceR^{\mi}$, output $y_i\in\spaceY_i\subset\spaceR^{h_i}$, and the state disturbance $w_i\in  \spaceR^{\ni}$ an independent, identically distributed noise sequence with distribution $w_i\sim \mathbb{P}_\ruohan{{w_i}}(\cdot)$, i.e., $w_i[t_1]$ and $w_i[t_2]$ are independent for all $t_1\neq t_2$. Each agent $\mdpM^i$ \eqref{eq:decoupled_hd_mdp} is initialized with state $x_i[0]$. 

For the combined multi-agent system $\M$, the collective state is given as $x=\operatorname{col}(x_1,\dots,x_N)\in\mathbb{R}^{n}, n=\sum_{i=1}^N n_i,$
together with the collective stochastic dynamics 
\begin{align}\label{eq:composed_model}
		\left\{\begin{array}{ll}	x^+
		  &=    
			f(x,u) + 
			w \\
			y
		 &=  
			\ruohan{g}(x)  \end{array}\right. 
\end{align}
where $u=\operatorname{col}(u_1 \dots u_N)\in\spaceR^{m}$ with $m=\sum_{i=1}^N \mi$, and $y=\operatorname{col}(y_1 \dots y_N)\in\spaceR^{h}$ with $h=\sum_{i=1}^N h_i$.
The state disturbance $w\in \mathbb{R}^{n}$ is distributed as $w \sim \mathbb{P}_{w}(\cdot)$ with $\mathbb{P}_w:=\prod_{i=1}^n \mathbb{P}_{w_i}$ the joint distribution. The random variables $w_i$ and $w_{j}$ are independent for all $i\neq j$, that is, the stochastic disturbances of the multi-agent system are independent. The multi-agent system \eqref{eq:composed_model} is initialized as $x[0]:= \operatorname{col}(x_1[0]\ldots x_N[0])$ by construction.
\par
The \textit{execution (state trajectory)} of the system \eqref{eq:composed_model} is a sequence of states $\boldsymbol{x}_{[0,T]}:=\{x[t] | t\!=\!0,\!\ldots,\!T\}$
initialized with  $x[0]$. 
Consecutive states $x[t+1]$ are obtained from realizations $x[t+1]\sim \Tr(\cdot | x[t],u[t])$ of the controlled Borel-measurable stochastic kernel, which is defined for system \eqref{eq:composed_model} as $\Tr(dx[t+1] | x[t],u[t]):=\mathbb{P}_w(dx[t+1]-f(x[t],u[t]))$. Meanwhile, the \textit{output trajectory} of the system \eqref{eq:composed_model} is a sequence of outputs $\boldsymbol{y}_{[0,T]}:=\{y[t]|t\!=\!0,\ldots,T\}$,
where $y[t]$ is obtained as $y[t]=\ruohan{g}(x[t])$.
\ruohan{
\par When the control inputs are selected only based on the current states, this is referred to as a Markov policy.
\begin{definition}[Markov policy $\boldsymbol{\pi}$] \label{def:markov_policy} 
     A Markov policy $\boldsymbol{\pi}$ is a sequence $ \boldsymbol{\pi}:= \{ \pi[t] | t=0,1,\ldots \}$ of maps $\pi[t]: \spaceX \rightarrow \spaceU   $ that assign an action $u\in \spaceU$ to each state $x\in\spaceX$.   
\end{definition}
More general control strategies may depend on the full history. In this paper, we restrict to finite-memory control strategies depending on the specification, the definition of which follows in the next subsection. }


To each agent $\mdpM^i$, we associate a set of unique atomic propositions $\AP_i:=\left\{p_1,\dots,p_{|\AP_i|}\right\}$ for which 
	$\AP_i\cap\AP_j=\emptyset,\textmd{ if }i \neq j,\forall i,j \in \{1,\ldots,N\}$. A set of atomic propositions $\AP_i$ defines the alphabet $\Sigma^i:=2^{\AP_i}$. For each output $y_i$, the atomic propositions can either be true or false. The set of true propositions for a given output is defined by the labeling map $L^i:\spaceY_i\rightarrow 2^{\AP_i}$. 
\par The output trajectory $\boldsymbol{y}=y[0],y[1],\ldots$ of system \eqref{eq:composed_model} is mapped to the word $\word:=l_0,l_1,\ldots$ using the labelling map $L:\spaceY\rightarrow 2^\AP$ defined as $L\bigl( \operatorname{col}(y_1 \dots y_N) \bigr):= \bigcup_{i=1}^N L^i(y_i)\in 2^{\AP}$,
that translates each output to a specific letter $l_t=L(y[t])$. Similarly, output suffixes $\boldsymbol{y}_t:=y[t],y[t+1],\ldots$ are translated to word suffixes $\word_t:=l_t,l_{t+1},\ldots$. 

\begin{figure}[htp]
    \centering
    \includegraphics[width=0.35\textwidth]{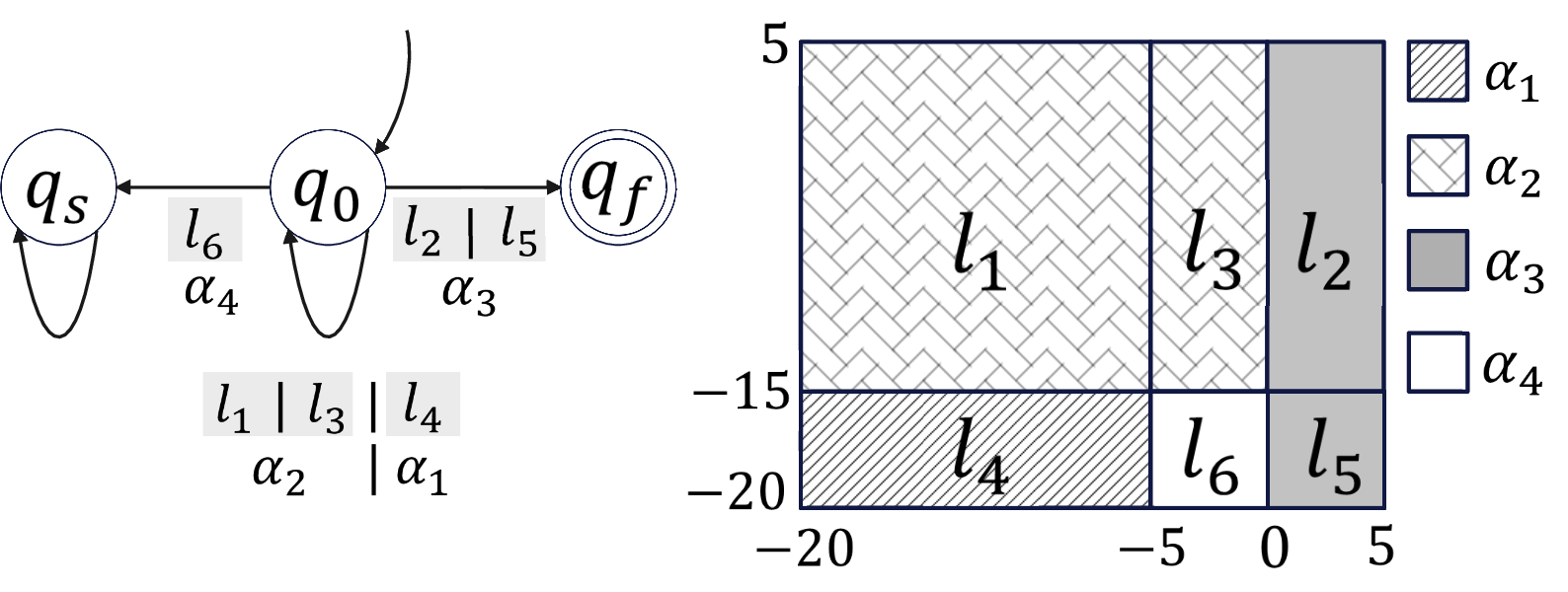} \vspace{-0.2cm}
    \caption{\begin{small}Left: Letter- or Boolean-formula-triggered transitions of DFA ($q_s$ denotes $q_{\text{sink}}$ for simplicity), right: labeled output space. 
    \end{small}}
    \label{fig:dfa_double_labeled_region}
    \vspace{-0.2cm}
\end{figure} 

In this paper, we are interested in temporal specifications $\phi$, e.g., syntactically co-safe linear temporal logic (\textit{scLTL}) \cite{kupferman2001model} and linear temporal logic on finite traces (\textit{LTL\textsubscript{f}}) \cite{de2013linear}, that can be automatically translated to a deterministic finite automaton (DFA) via off-the-shelf tools. In the sequel, we drop the subscript $\phi$ for the sake of simple presentation. We give the definition of Deterministic Finite Automaton (DFA) as follows:

\begin{definition}[DFA]
\label{def:DFA} \hspace{1mm}
    A deterministic finite automata (DFA) is defined by the tuple $\DFA=(Q,q_0,\Sigma_\DFA,\tau_{\DFA},q_f)$.
Here, $Q$, $q_0$, and $q_f$ denote the set of states, initial state, and accepting state, respectively. Furthermore, $\Sigma_\DFA=2^{\AP}$ denotes the input alphabet and $\tau_\DFA:Q\times \Sigma_\DFA \rightarrow Q$ is a transition function.
\end{definition}
We denote a DFA translated from a specification $\phi$ as $\DFA_{\phi}$.

\noindent{\bfseries Boolean-formula-labelled transitions of DFA.}\hspace{1mm}Given a DFA $\DFA=(Q,q_0,\Sigma_\DFA,\tau_{\DFA},q_f)$, let us define a set $\subphiSet:=\{\subphi\}$ composed of Boolean formulae $\subphi$, that allows the rewriting of DFA transition as $\tau_{\DFA}: Q\times \subphiSet \rightarrow Q$. Each $\subphi$ is a conjunction of (possibly negated) atomic propositions: $
    \subphi::=  p|  \neg p | \alpha _1\wedge\alpha_2.
$
Furthermore, for every letter $l\in \Sigma_\DFA$ there exists exactly one such $\subphi$ with $l \models \subphi$. See Fig.~\ref{fig:dfa_double_labeled_region} for an intuitive grasp on $\subphi$. To agent $i$, we define $\subphi_i$ as $\subphi_i:=\{ \bigwedge_{l \vDash \subphi} l^i  \}$, with $l^i:=L^i(y_i)$.

\subsection{Specification}\vspace{-0.2cm}

\colorlet{BluePanelStroke}{blue!45}
\colorlet{BluePanelFill}{blue!10}
\colorlet{RedPanelStroke}{red!45}
\colorlet{GreenPanelTone}{green!70}

\def\N{5}
\def\s{0.4}

\tikzset{
  tensornode/.pic={
    \node at (0,0) {$\otimes$};
    \foreach \i in {1,...,\N} {
      \pgfmathsetmacro{\val}{int(30+10*\i)}
      \fill[red!\val] (\i*\s,0) rectangle ++(\s,\s);
    }
    \foreach \i in {1,...,\N} {
      \pgfmathsetmacro{\val}{int(20+10*\i)}
      \fill[green!\val] (0,\i*\s) rectangle ++(\s,\s);
    }
    \foreach \i in {1,...,\N} {
      \pgfmathsetmacro{\val}{min(max(10+20*\i,0),100)}
      \fill[blue!\val] (0,0,\i*\s) rectangle ++(\s,\s);
    }
  }
}

\tikzset{
  policytensorglyph/.pic={
    \foreach \i in {1,...,\N} {
      \pgfmathsetmacro{\val}{min(max(10+20*\i,0),100)}
      \fill[blue!\val] (0,0,\i*\s) rectangle ++(\s,\s);
    }
    \fill[GreenPanelTone] ({-0.90*\s},{(\N-4.20)*\s}) rectangle ++(\s,\s);
    \fill[RedPanelStroke] ({0.55*\s},{-1.35*\s}) rectangle ++(\s,\s);
  }
}

\tikzset{
  policyexample/.pic={
    \begin{scope}[shift={(0.35,0.3)}, scale=1.2]
      \pic{policytensorglyph};
    \end{scope}
    \draw[-{Stealth[length=2.2mm,width=1.5mm]}, thick, dotted, black]
      (1.1,0.15) -- (3.55,0.15);
    \node[above=2pt, font=\large] at (2.35,0.15) {$\pi_{q}^{(i)}(\hat{x}_{i,\bullet})$};
    \matrix (m) [matrix of nodes,
      ampersand replacement=\&,
      nodes={draw, minimum width=0.78cm, minimum height=0.55cm,
        align=center, font=\normalsize, anchor=center,
        text height=2.2ex, text depth=0.8ex},
      row sep=-\pgflinewidth, column sep=-\pgflinewidth, anchor=west
    ] at (3.85,0.05) {
      $\hat{u}_i^1$ \& $\hat{u}_i^2$ \& $\hat{u}_i^3$ \& $\hat{u}_i^4$ \& $\hat{u}_i^5$ \& \celldots \\
      0.7 \& 0.8 \& 0.9 \& 0.8 \& 0.7 \& \celldots \\
    };
    \begin{scope}[on background layer]
      \node[fit=(m-1-3)(m-2-3), fill=yellow, opacity=0.45,
            inner sep=0pt, rounded corners=1pt] {};
    \end{scope}
  }
}

\begin{figure*}[bt]
  \centering
  \resizebox{0.75\textwidth}{!}{%
\begin{tikzpicture}[
  transform shape, >=Stealth,
  arrow/.style={-{Stealth[length=3mm,width=2mm]}, thick, shorten >=2pt, shorten <=2pt},
  thinlink/.style={-Stealth, semithick, shorten >=1pt, shorten <=1pt},
  model/.style={draw, thick, minimum size=0.6cm, align=center,
                font=\bfseries\footnotesize, rounded corners},
  every node/.style={font=\scriptsize}
]
\tikzset{abslabel/.style={midway, above=2pt, inner sep=1pt}}

\def\colgap{2.8cm}
\def\rowgap{1.3cm}
\def\mergeX{1.9cm}
\def\gridW{1cm}
\def\gridH{1cm}
\def\hdrRaise{1.5mm}
\def\hdrSide{0.7cm}
\def\fcBoxW{2.4cm}
\def\fcBoxH{0.80cm}
\def\fcVgap{2.2cm}
\newlength{\splitShift}
\setlength{\splitShift}{3mm}


\node[draw, thick, minimum width=\gridW, minimum height=\gridH, rounded corners,
   ] (cont1) {};
\node[model, above=\hdrRaise of cont1,   label=above:{\textbf{Continuous gMDPs}}] (M1) {$\mdpM^1$};

\node[ above=\hdrRaise of M1] (M1above) {};

\node[left=\hdrSide of M1]  (u1) {$u_1$};
\node[right=\hdrSide of M1] (x1) {$x_1$};
\draw[arrow] (u1)--(M1.west); \draw[arrow] (M1.east)--(x1);
\begin{scope}[shift={(cont1.center)}]
  \foreach \r/\op in {0.42/18, 0.30/32, 0.18/52, 0.08/78}
    \fill[black!\op!white, opacity=0.75] (0.04,0) ellipse ({\r*\gridW} and {\r*0.80*\gridH});
  \foreach \r in {0.42, 0.30, 0.18}
    \draw[black!65, semithick] (0.04,0) ellipse ({\r*\gridW} and {\r*0.80*\gridH});
  \fill[black!85] (0.04,0) circle (1.6pt);
\end{scope}

\node[draw, thick, minimum width=\gridW, minimum height=\gridH, rounded corners,
      below=\rowgap of cont1] (cont2) {};
\node[model, above=\hdrRaise of cont2] (M2) {$\mdpM^2$};
\node[left=\hdrSide of M2]  (u2) {$u_2$};
\node[right=\hdrSide of M2] (x2) {$x_2$};
\draw[arrow] (u2)--(M2.west); \draw[arrow] (M2.east)--(x2);
\begin{scope}[shift={(cont2.center)}, rotate=-22]
  \foreach \r/\op in {0.42/18, 0.30/32, 0.18/52, 0.08/78}
    \fill[black!\op!white, opacity=0.75] (-0.04,0) ellipse ({\r*1.22*\gridW} and {\r*0.58*\gridH});
  \foreach \r in {0.42, 0.30, 0.18}
    \draw[black!65, semithick] (-0.04,0) ellipse ({\r*1.22*\gridW} and {\r*0.58*\gridH});
  \fill[black!85] (-0.04,0) circle (1.6pt);
\end{scope}

\node[draw, thick, minimum width=\gridW, minimum height=\gridH, rounded corners,
      below=\rowgap of cont2] (cont3) {};
\node[model, above=\hdrRaise of cont3] (M3) {$\mdpM^3$};
\node[left=\hdrSide of M3]  (u3) {$u_3$};
\node[right=\hdrSide of M3] (x3) {$x_3$};
\draw[arrow] (u3)--(M3.west); \draw[arrow] (M3.east)--(x3);
\begin{scope}[shift={(cont3.center)}, rotate=18]
  \foreach \r/\op in {0.42/18, 0.30/32, 0.18/52, 0.08/78}
    \fill[black!\op!white, opacity=0.75] (0,-0.04) ellipse ({\r*0.68*\gridW} and {\r*1.12*\gridH});
  \foreach \r in {0.42, 0.30, 0.18}
    \draw[black!65, semithick] (0,-0.04) ellipse ({\r*0.68*\gridW} and {\r*1.12*\gridH});
  \fill[black!85] (0,-0.04) circle (1.6pt);
\end{scope}


\node[draw, thick, minimum width=\gridW, minimum height=\gridH, rounded corners,
      right=\colgap of cont1] (grid1) {};
\node[model, above=\hdrRaise of grid1,,
label=above:{\textbf{Abstract Finite-State gMDPs}}] (Mh1) {$\hat{\mdpM}^1$};
\node[left=\hdrSide of Mh1]  (ua1) {$\hat{u}_1$};
\node[right=\hdrSide of Mh1] (xs1) {$\hat{y}_1$};
\draw[arrow] (ua1)--(Mh1.west); \draw[arrow] (Mh1.east)--(xs1);
\foreach \k in {1,2,3}{
  \draw ($(grid1.south west)+(\k*\gridW/4,0)$) -- ++(0,\gridH);
  \draw ($(grid1.south west)+(0,\k*\gridH/4)$) -- ++(\gridW,0); }

\node[draw, thick, minimum width=\gridW, minimum height=\gridH, rounded corners,
      right=\colgap of cont2] (grid2) {};
\node[model, above=\hdrRaise of grid2] (Mh2) {$\hat{\mdpM}^2$};
\node[left=\hdrSide of Mh2]  (ua2) {$\hat{u}_2$};
\node[right=\hdrSide of Mh2] (xs2) {$\hat{y}_2$};
\draw[arrow] (ua2)--(Mh2.west); \draw[arrow] (Mh2.east)--(xs2);
\foreach \k in {1,2,3}{
  \draw ($(grid2.south west)+(\k*\gridW/4,0)$) -- ++(0,\gridH);
  \draw ($(grid2.south west)+(0,\k*\gridH/4)$) -- ++(\gridW,0); }

\node[draw, thick, minimum width=\gridW, minimum height=\gridH, rounded corners,
      right=\colgap of cont3] (grid3) {};
\node[model, above=\hdrRaise of grid3] (Mh3) {$\hat{\mdpM}^3$};
\node[left=\hdrSide of Mh3]  (ua3) {$\hat{u}_3$};
\node[right=\hdrSide of Mh3] (xs3) {$\hat{y}_3$};
\draw[arrow] (ua3)--(Mh3.west); \draw[arrow] (Mh3.east)--(xs3);
\foreach \k in {1,2,3}{
  \draw ($(grid3.south west)+(\k*\gridW/4,0)$) -- ++(0,\gridH);
  \draw ($(grid3.south west)+(0,\k*\gridH/4)$) -- ++(\gridW,0); }

\draw[arrow] (cont1.east) -- node[abslabel]{Abstraction} (grid1.west);
\draw[arrow] (cont2.east) -- node[abslabel]{Abstraction} (grid2.west);
\draw[arrow] (cont3.east) -- node[abslabel]{Abstraction} (grid3.west);

\node[circle, minimum size=2pt, inner sep=0pt, fill=black,
      right=\mergeX of grid2.east] (merge) {};
\draw[thinlink] (grid1.east) -- ++(\mergeX-0.4cm,0) |- (merge);
\draw[thinlink] (grid2.east) -- (merge);
\draw[thinlink] (grid3.east) -- ++(\mergeX-0.4cm,0) |- (merge);


\node[anchor=center, font=\bfseries\normalsize]
  (fcVI) at ($(merge) + (4cm, 4.25cm)$) {};

\node[draw, thick, rounded corners=4pt,
      minimum width=5cm, minimum height=\fcBoxH,
      align=center, font=\bfseries\small,
      below=\fcVgap of fcVI]
  (fcBell) {
 \begin{tikzpicture}
 	\begin{axis}[width=2.3cm, height=1.65cm, scale only axis,
 		axis lines=none, ticks=none, enlargelimits=false,
 		view={-35}{35}, z buffer=sort, shader=interp,
 		colormap={rgbmap}{
 			rgb255(0pt)=(30,30,220);    
 			rgb255(33pt)=(30,180,30);   
 			rgb255(66pt)=(220,30,30)    
 		}]
 		\addplot3[xshift=-0.4cm, surf, domain=0:1, y domain=0:1, samples=22, samples y=22]{
 			0.9*exp(-7*((x-0.25)^2+(y-0.35)^2))
 			+0.7*exp(-10*((x-0.75)^2+(y-0.65)^2))
 			+0.15*sin(6*deg(x))*cos(6*deg(y))};
 	\end{axis}
 	\end{tikzpicture}\\
  Update value function $\mathcal V^\pi$
};

\node[draw, thick, rounded corners=4pt,
minimum width=\fcBoxW, minimum height=\fcBoxH,
align=center, font=\bfseries\small,
below=.5cm of fcBell, xshift=-.8cm,  label=above:{\textbf{+}}]
(fcCorA) {Error Correction ($\delta$)\\Thm.~\ref{theorem:satprob_lb_robusttree}};

\node[draw, thick, rounded corners=4pt,
minimum width=\fcBoxW, minimum height=\fcBoxH,
align=center, font=\bfseries\small,
below=1.8cm of fcBell,xshift=1.2cm]
(fcCorB) {Aposteriori Correction ($\delta$)\\
Thm.~\ref{theorem:branch_length_lossProb}};

\node[
align=center, font=\bfseries\small,
right=.75cm of fcCorB]
(sol) {\underline{Guaranteed Satisfaction probability}};
]

\node[draw, thick, rounded corners=4pt,
      minimum width=\fcBoxW, minimum height=\fcBoxH,
      align=center, font=\bfseries\small,
      above=1.2cm of fcBell]
  (fcPol) {Policy update $\pi \leftarrow \pi'$};


\draw[arrow]
  (fcPol.south) -- node[right=2pt, font=\small, yshift=-10pt]{updated $\pi$}
  (fcBell.north);

\draw[arrow]
  (fcBell.west) -- ++(-1.20cm,0)
  |- node[right=1pt, font=\small, pos=0.25]{updated $\mathcal V^\pi$} (fcPol.west);

\draw[arrow, dashed]
(fcBell.south) ++(1.20cm,0) -- node[right=2pt, font=\small, yshift=-10pt]{Converged $\mathcal V^\pi$}
(fcCorB.north);

\draw[arrow]
(fcCorB.east) --  
(sol.west);


\coordinate (fcMid) at ($(fcVI.north)!0.50!(fcCorA.south)$);

\node[anchor=west, inner sep=6pt] (treeInner)
  at ($(fcVI.east |- fcMid) + (3cm, 0)$)
{%
  \begin{tikzpicture}[scale=0.60, transform shape,
    grow=down, level distance=2.0cm,
    edge from parent/.style={draw, thick, dashed, dash pattern=on 3pt off 2pt,
      -{Stealth[length=3mm,width=2mm]}},
    every node/.style={inner sep=0pt},
    level 1/.style={sibling distance=5.5cm},
    level 2/.style={sibling distance=2.75cm},
    level 3/.style={sibling distance=1.5cm}]
    \tikzset{tn/.style={minimum size=1cm, scale=0.4}}
    \node[tn]{\tikz\pic{tensornode};}
      child{ node[tn]{\tikz\pic{tensornode};}
        child{ node[tn]{\tikz\pic{tensornode};} }
        child{ node[tn]{\tikz\pic{tensornode};} }
      }
      child{ node[tn]{\tikz\pic{tensornode};}
        child{ node[tn]{\tikz\pic{tensornode};} }
        child{ node[tn]{\tikz\pic{tensornode};} }
      };
  \end{tikzpicture}%
};

\node[draw, thick, rounded corners=5pt, fit=(treeInner),
      inner xsep=10pt, inner ysep=8pt,
      label={[font=\small\bfseries]below:{Tensor tree representation}}]
  (treeFrame) {};
\node[anchor=south, font=\small\bfseries, yshift=2pt]
  at (treeFrame.north) {Subsection.~\ref{sec:subsec:cdc}};

\draw[blue!70,  dashed]
(fcBell.north east) -- (treeFrame.north west);
\draw[blue!70, dashed]
(fcBell.south east) -- (treeFrame.south west);

\coordinate (xs3R) at ($(xs3.east)+(\splitShift,0)$);

\begin{pgfonlayer}{background}
  \node[fit=(u1.west)(xs3R)(M1.north)(M1above)(cont3.south),
        fill=red!10, draw=red!45, rounded corners=6pt,
        inner xsep=10pt, inner ysep=10pt] (bgLeft) {};
  \node[anchor=north, font=\footnotesize\bfseries, text=black]
    at ($(bgLeft.south)+(0,-3pt)$) {Sec.~3: Abstraction};

  \coordinate (rightTop) at ($(fcPol.north -| treeFrame.east) + (10pt, 4pt)$);
  \coordinate (leftBot) at ($(sol.south -| merge) + (12pt,-1pt)$);
  \node[fit=(rightTop)(leftBot),
        fill=blue!10, draw=blue!45, rounded corners=6pt,
        inner xsep=10pt, inner ysep=10pt] (bgRight) {};
  \node[anchor=north, font=\footnotesize\bfseries, align=center, text=black]
    at ($(bgRight.south)+(0,-3pt)$) {Sec.~\ref{sec:tensor_tree_ap} $\&$ Sec.~\ref{sec:new_robust_tensor_tree}: Robust Value Iteration with Tensors};
\end{pgfonlayer}

\end{tikzpicture}
} 
  \caption{\begin{small}\ruohan{Control-synthesis for a three-agent system. Each agent $\mdpM^i$ \eqref{eq:decoupled_hd_mdp} is abstracted as a gMDP $\hat{\mdpM}^i$ (red-shaded block; Sec.~\ref{sec:simrel}). The per-agent value components are then merged in every tree node and propagated through the tensor tree (blue-shaded block; Sec.~\ref{sec:tensor_tree_ap} and Sec.~\ref{sec:new_robust_tensor_tree}), yielding a high-dimensional satisfaction probability.}\end{small} }
  \label{fig:paper_overview}
\end{figure*}

For a given word $\word= l_0,l_1,\ldots$, a run of a DFA $\DFA$ defines a trajectory $
q_0,q_1,\ldots 
$
that starts with $q_0$ and evolves according to
$
q_{t+1} = \tau_{\DFA}(q_t,\subphi_t). 
$
The word $\word_{[0,T]} $ is accepted by  $\DFA_\phi$ if $\exists t\in[0,T]$ such that $q_t =q_f$ for the corresponding trajectory $q_0,q_1,\ldots,q_T$. This is denoted as $\word_{[0,T]} \models \phi$. For $T\rightarrow\infty$, we denote  $\word  \models \phi$. Given a specification $\phi$ and its DFA $\DFA=(Q,q_0,\Sigma_\DFA,\tau_\DFA,q_f)$, we label the output trajectory $\boldsymbol{y}$ of a controlled system $\mdpM\times \mathbf{C}_\pi$ and obtain a word $\word$. The controlled system satisfying the specification can be translated to an acceptance problem, i.e., the probability that the word $\word$ is accepted; \ruohan{or equivalently as a reachability problem on the product system $\mdpM\times \DFA_\phi$ \cite{tkachev2017quantitative,haesaert2020robust}.
\par
A finite-memory policy $\boldsymbol{\pi}$ is a sequence $ \boldsymbol{\pi}:= \{ \pi[t] | t=0,1,\ldots \}$ of maps $\pi[t]: \spaceX \times Q \rightarrow \spaceU   $ that assign an action $u\in \spaceU$ to each state pair $(x,q)\in \spaceX \times Q$.}   



\subsection{Problem statement $\&$ Approach}\vspace{-0.2cm}

Correct-by-design methods for infinite-state (particularly continuous-state) stochastic systems typically rely on finite-state abstractions \cite{tabuada2009verification} and on approximate simulation relations \cite{haesaert2020robust,schon2023verifying} to carry guarantees across models. Our prior work \cite{wang2025unraveling} mitigated the curse of dimensionality in finite-state control synthesis while preserving provable guarantees. 
In this paper, we focus on continuous-state systems and develop an approach that preserves correct-by-design guarantees while remaining computationally and memory efficient as state spaces grow. \ruohan{We consider multi-agent systems with decoupled agent dynamics and a coupled specification.  
}
\par
\noindent 
\textbf{\textit{Problem statement:}}\hspace{1mm}\textit{Given a specification $\phi$ with DFA $\DFA_\phi$ and a probability $p\in[0,1]$, synthesize control strategies $\mathbf{C}_{\boldsymbol{\pi}}$ for a multi-agent system $\mdpM$ such that the controlled system satisfies $\phi$ with a lower bound, i.e.,
\begin{tightdisplay}
\begin{align}
    \ruohan{\mathbb{P}^{x_0}_{\!\mdpM \times \mathbf{C}_{\boldsymbol{\pi}}}(\word\models \phi) \geq p}.\label{eq:pb_statement}
\end{align}
\end{tightdisplay}
}
Fig.~\ref{fig:paper_overview} presents a schematic overview of our approach, exemplified on a three-agent system. 

\vspace{-0.1cm} 


\section{Robust dynamic programming for multi-agent systems}\label{sec:simrel}
A stochastic difference equation can also be modelled as a general Markov decision process defined as follows.
\begin{definition}
[general Markov decision processes (gMDP)]\label{def:gmdp}\hspace{2mm} A  gMDP is denoted as $\mathbf M:=(\spaceX,\spaceU,\spaceY, \Tr,x_0,\ruohan{g} )$
	with   
	\begin{itemize}
		\item 
		a Polish state space $\spaceX$ with states $x\in \spaceX$;
		\item a Polish input space $\spaceU$ with inputs/actions $u\in\spaceU$;
		\item an output space $\spaceY$ decorated with metric $\mathbf{d}_{\spaceY}$;
		\item a stochastic kernel $\Tr :\spaceX \times \spaceU \times \mathcal{B}(\spaceX) \rightarrow  [0,1]$, that assigns to each state-action pair $x\in\spaceX$ and $u\in\spaceU$ a probability measure $\Tr(\cdot \mid x,u)$ over $(\spaceX,\mathcal{B}(\spaceX))$;
		\item an initial state $x_0\in\spaceX$;
		\item a measurable output map $\ruohan{g}:\spaceX \rightarrow \spaceY$.
	\end{itemize} 
\end{definition}

We say that a gMDP is finite if $\spaceX$, $\spaceY$, and $\spaceU$ are finite. In this paper, each agent $i$ \eqref{eq:decoupled_hd_mdp} can be written as a gMDP 
\begin{align}
    \mdpM^i=(\spaceX_i,\spaceU_i,\spaceY_i, \Tr_i,x_{i,0},\ruohan{g_i} ), \quad  i = 1,\ldots,N, \label{eq:agent_gmdp}
\end{align}
where $\Tr_i(x_{i,t+1}|x_{i,t},u_{i,t})=\mathbb{P}_{w_i}(x_{i,t+1}-f(x_{i,t},u_{i,t}))$. Similarly, the associated multi-agent gMDP can be modeled \eqref{eq:composed_model} via Cartesian composition:
\begin{align} \label{eq:cart_gmdp}
    \mdpM=(\spaceX,\spaceU,\spaceY,\Tr,x_0,\ruohan{g} )
\end{align}
with $\spaceX=\prod_{i=1}^{N}\spaceX_i$, $\spaceU=\prod_{i=1}^{N}\spaceU_i$, $\spaceY=\prod_{i=1}^{N}\spaceY_i$, $\Tr=\prod_{i=1}^{N}\Tr^i$, $x_0:=\operatorname{col}(x_{1,0},...,x_{N,0})$, and $\ruohan{g=\operatorname{col}(g_{1},...,g_{N})}$.


\vskip-.2cm
\subsection{Dynamic programming}\vskip-.2cm
For a multi-agent system $\mdpM$ and a specification $\phi$ with its DFA $\DFA$, we 
define value functions $\tensorV_{q,T}^{\boldsymbol{\pi}}:\spaceX  \rightarrow [0,1]$ as the probability that the generated word $\word_{[0,T]}$ gets accepted \cite{wang2025unraveling,bertsekas1996stochastic}.
Given a Markov policy $\boldsymbol{\pi}:= \{ \pi[t] | t=0,1,\ldots,T-1\}$, value functions can iteratively be computed starting from  $\tensorV_{q,0} = 1$ if $q=q_f$, and $0$, otherwise and iterated as
\begin{equation}\label{eq:vi_coupled_similar_model}
	\begin{aligned}
		\!\!&\tensorV_{q,t+1}^{\boldsymbol{\pi}}(x)= \indicator_{q_f}(q)
 +\indicator_{Q\setminus\{q_f\}}(q)  
		\expectation^{\pi{[T\! -1- t]}}_{x,q}[\tensorV_{q',t}^{\boldsymbol{\pi}}(x') ] \!\!
	\end{aligned}
\end{equation}where \ruohan{$\expectation_{x,q}^{\pi}$ is the expectation conditioned on $\pi$ and over probabilistic transitions of $x$ and $q$, and defined as $\expectation^{\pi}_{x,q}[\tensorV_{q',t}^{\boldsymbol{\pi}}(x')]:= \expectation[\tensorV_{q',t}^{\boldsymbol{\pi}}(x') | (x,q),u=\pi(x,q)] $ with $q':=\tau(q,\subphi)$ with $\subphi=L(\ruohan{g}(x'))$.} $\indicator_A(a)$ is an indicator function, defined as $\indicator_A(a):=1$ if $a\in A$, and $0$, otherwise.
As in \cite{wang2025unraveling,haesaert2020robust}, the subsequent proposition follows. 
\begin{proposition}[Specification satisfaction.] \label{prop:spec_satisfaction}
\hspace{1mm}For a given policy $\boldsymbol{\pi}=\{\pi[t]|t=0,1,\ldots,T-1\}$, 
     the satisfaction probability within time horizon $T$ for specification $\phi$ of the multi-agent system $\mathbf{M}$ is defined as 
\begin{equation}
\begin{aligned}
\ruohan{\mathbb{P}^{x_0}_{\mdpM \times \boldsymbol{\pi}}(\word_{[0,T]} \models \phi ):=\tensorV_{\bar{q}_0,T}^{\boldsymbol{\pi}}(x_0) }
\end{aligned}
\end{equation} with $\bar q_0= \tau_{\mathcal A}(q_0,L(\ruohan{g}(x_0)))$,   and 
computed as in \eqref{eq:vi_coupled_similar_model}.
\end{proposition}

\noindent {\bfseries A general form of DFA-informed operators.} Value functions $\tensorV_{q,T}^{\boldsymbol{\pi}}(x)$ can be recursively computed as
\begin{align} \label{eq:Viterations}
\tensorV_{q,t+1}^{\boldsymbol{\pi}}(x)=\sum_{(\subphi,q')\in {\small \text{Succ}}(q)}\op_{\subphi}^{\pi_q[T-1-t]}(\tensorV^{\boldsymbol{\pi}}_{q',t})(x)
\end{align}for all $q\in Q\setminus\{q_f\}$, where 
$t=0,\ldots,T-1$, ${\small \text{Succ}}(q)$ is defined for DFA as ${\small \text{Succ}}(q):=\{(\subphi,q')| q'=\tau_\DFA(q,\subphi)\}$, and $\op_{\subphi}^{\pi_q}$ is the DFA-informed operator originally proposed in \cite{wang2025unraveling}, defined as
\begin{align}\label{eq:op_general}
\op_{\subphi}^{\pi}(\tensorV)(x):=	\expectation^{\pi}_{x}[\mathcal{L}_{\subphi}(\ruohan{g}(x'))\tensorV(x')],
\end{align}
with 
the indicator functions $\mathcal{L}_{\subphi}:\spaceY\rightarrow \{0,1\}$ defined for each formula $\subphi$ as
\begin{equation} \label{eq:indicator_labeling}
    \mathcal{L}_{\subphi}(y')=\left\{\begin{array}{l}
1, \quad \text{if }L(y') \vDash \subphi\\
0, \quad \text{otherwise.}
\end{array}\right.
\end{equation}
Dynamic programming is generally intractable on gMDPs with infinite state spaces, particularly uncountable ones (e.g., Euclidean state spaces), yet feasible on finite gMDPs. In this paper, we approximate the continuous-state multi-agent system using a finite-state model and perform value iteration on it. To soundly refine the synthesized controller back to the continuous-state system, we quantify the abstraction error via approximate simulation relations, as introduced in the next section.

\subsection{Approximate simulation relations} \label{subsec:asr}
\par
Consider a \textbf{concrete} system modeled as a gMDP $$\mdpM=(\spaceX,\spaceU,\spaceY,\Tr,x_0,\ruohan{g})$$ and its \textbf{approximation} $$\hat{\mdpM}=(\hat{\spaceX},\hat{\spaceU},\hat{\spaceY},\hat{\Tr},\hat{x}_0,\ruohan{\hat{g}}).$$ 


To quantify the error caused by $\hat{\mdpM}$, we first quantify the similarity between probability distributions $\Tr$ and $\hat{\Tr}$. Let $\relation$ be a measurable relation $\relation \subset \hat{\spaceX} \times \spaceX$ that relates states of the concrete system with those of the approximation. To use this relation, we need to define its lifted version, which is denoted as $\bar{\relation}_{\delta}$ where $\delta$ quantifies the probability mass that falls outside of the lifting. More precisely we say that $(\Delta,\Theta)$ belongs to the $\delta$-lifted relation $\bar{\relation}_{\delta}$ if there exists a probability measure $\mathbb{W}$, referred to as a lifting, with probability space $(\hat{\spaceX}\times \spaceX,\mathcal{B}(\hat{\spaceX}\times \spaceX),\mathbb{W})$ such that
\begin{equation}
\begin{aligned}
    \textbf{L1.} \quad & \mathbb{W}(\hat{A}\times \spaceX)=\Delta(\hat{A}) \quad \forall \hat{A}\in \mathcal{B}(\hat{\spaceX}) ,\label{eq:delta_lift_L1}\nonumber
    \\
    \textbf{L2.} \quad & \mathbb{W}(\hat{\spaceX}\times A)=\Theta(A) \quad \forall A\in \mathcal{B}(\spaceX) ,\nonumber
    \\
    \textbf{L3.} \quad & \mathbb{W}(\relation) \leq 1-\delta.\nonumber
\end{aligned}
\end{equation}
If the pair of probability distribution belongs to the lifted relation $(\Delta,\Theta)\in \bar{\relation}_{\delta}$, then we denote this as $\Delta \bar{\relation}_{\delta} \Theta$. 
$\delta$ quantifies the amount of the probability distribution that cannot be coupled into the relation $\relation$. In the remainder, we will consider lifting of the stochastic kernels of $\hat \mdpM$ and $\M$, that is,  $\hat \Tr \bar{\relation}_\delta \Tr$. 
Consider an \textit{interface function} \cite{girard2007approximation,haesaert2020robust} that refines control actions as follows:
\begin{align}
    \mathcal{U}:\hat{\spaceU} \times \hat{\spaceX} \times \spaceX 
\rightarrow \ruohan{\mathcal{P}(\spaceU)}. \label{eq:interface_function}
\end{align}
Here, an interface function refines any control action $\hat{u}$ synthesized over the approximation $\hat{\mdpM}$ to an action that's applied to the concrete system $\mdpM$. This interface is used with $\delta$-lifting to define approximate stochastic simulation relations on gMDPs as defined in \cite{haesaert2020robust,schon2023verifying} and recalled next.
%
\begin{definition}
    [$(\epsilon,\delta)$-stochastic simulation relation] \label{def:simulation-relation} 
    \hspace{1mm}Consider a concrete system $\mdpM=(\spaceX,\spaceU,\spaceY,\Tr,x_0,\ruohan{g})$ and its approximation $\hat{\mdpM}=(\hat{\spaceX},\hat{\spaceU},\hat{\spaceY},\hat{\Tr},\hat{x}_0,\ruohan{\hat{g}})$. If there exist a measurable relation $\relation \subset \hat{\spaceX} \times \spaceX $, a measurable interface function $\mathcal{U} : \hat{\spaceU}  \times \hat{\spaceX}  \times \spaceX  \rightarrow \mathcal{P}(\spaceU )$, and a measurable stochastic kernel $\mathbb{W} _{\Tr }:\hat{\spaceU}  \times \hat{\spaceX}  \times \spaceX  \rightarrow \mathcal{P}(\hat{\spaceX} \times \spaceX )$ such that
    \begin{align*}
    \textbf{A1.} \quad & (\hat{x}_0 ,x_0 )\in \relation  ,\nonumber
    \\
    \textbf{A2.} \quad & \forall (\hat{x},x) \in \relation: \boldsymbol{d}_{\spaceY}(\hat{y},y) \leq \epsilon \text{ with } y=\ruohan{g}(x) \text{ and }\hat{y}=\ruohan{\hat{g}}(\hat{x})
    \nonumber
    \\
    \textbf{A3.} \quad & \hat\Tr (\cdot | \hat{x} ,\hat{u} )\relation _{\delta } \Tr (\cdot|x ,\mathcal{U} (\hat{u} ,\hat{x} ,x )) 
    \forall \hat{u} \in\hat{\spaceU} \hspace{2mm}  \forall (\hat{x} ,x )\in \relation  \nonumber
    \\
    &\quad \quad \text{with the lifted kernel } \mathbb{W} _{\Tr }(\cdot|\hat{u} ,\hat{x} ,x ) ,\nonumber
\end{align*}
\noindent then $\hat{\mdpM} $ is $(\epsilon,\delta) $-stochastically simulated by $\mdpM $. We denote this by   $\hat{\mdpM}  \preceq^{\delta }_\epsilon \mdpM $.
\end{definition}

\ruohan{
\begin{remark}
Related metric notions of approximate simulation for labeled Markov processes have been studied in \cite{desharnais2004metrics} and could be used to complement the coupling-based formulation adopted in this paper when a more explicit
topological treatment is desired.
\end{remark}
}
\noindent {\bfseries 
	Composable simulations for multi-agent systems.}
Consider agent $i$ modeled as a gMDP $\mdpM^i$ in \eqref{eq:agent_gmdp} with state space $\spaceX_i$. Let $\hat{\mdpM}^i$ be its approximation with a measurable relation $\relation_i \subset \hat{\spaceX}_i\times\spaceX_i$ that yields a similarity quantification  
\begin{align}
    \hat{\mdpM}^i \preceq^{\delta_i}_{\epsilon_i} \mdpM^i \nonumber
\end{align}
with respect to metric $d_{\spaceY_i}(\hat{y}_i,y_i):=\norm{\hat{y}_i-y_i}$. We immediately compose the agent approximations to obtain $\hat{\mdpM}$ with state space $\hat{\spaceX}:=\prod_{i=1}^N \hat{\spaceX}_i$  that approximates the concrete multi-agent system $\M$ in \eqref{eq:cart_gmdp}. Then 
there exists a multi-agent-level relation $\relation\subset \hat{\spaceX} \times \spaceX$ induced by agent-wise relations $\relation_i$ as $\relation:=\{(\hat x, x)\in \hat{\mathbb{X}}\times \mathbb{X}| (\hat x_i,x_i)\in \relation_i,\forall i\}$ for which 
\begin{align} \label{eq:1-prod_delta}
    \hat{\mdpM}\preceq^{\boldsymbol{\delta}}_{\boldsymbol{\epsilon}} \mdpM, \quad \boldsymbol{\delta}:=1-\prod_{i=1}^{N}(1-\delta^i), \quad \boldsymbol{\epsilon}:=\max_i(\epsilon^i),
\end{align}
with respect to the distance metric defined as
\begin{align} \label{eq:max_distance_metric}
d_{\mathbb Y}(\hat y,y):=\max_i (d_{\mathbb Y_i}(\hat{y}_i,y_i)). 
\end{align}
$\boldsymbol{\delta}$ quantifies the amount of probability distribution that cannot be coupled into the relation $\relation$. 
\par We proceed to address output deviation $\epsilon$ by modifying indicator functions $\mathcal{L}_\subphi$ in \eqref{eq:indicator_labeling}. Similar existing approaches include signal-space robustness degree \cite{fainekos2009robustness} and approximation metrics for transition systems \cite{girard2007approximation}. For probabilistic models, metrics on labelled Markov processes quantify behavioural closeness, with zero distance yielding probabilistic bisimulation \cite{desharnais2004metrics}. In contrast, we define $\epsilon$-robust indicator functions by tightening labeling via contracting (and, when needed, inflation) of predicate regions. The same robust-labeling viewpoint is also used in \cite{liu2016finite}, Section~3. 
\par
\noindent {\bfseries $\epsilon$-robust indicator functions.}\hspace{1mm}Given an $\epsilon\geq 0$, we define the $\epsilon$-robust indicator function $\mathcal{L}_{\subphi}^{\epsilon}:\spaceY\rightarrow \{0,1\}$ as
\begin{tightdisplay}
\begin{equation} \label{eq:robust_indicator}
    \mathcal{L}_{\subphi}^{\epsilon}(y)=\left\{\begin{array}{l}
1, \quad \text{if }L(y)\vDash \subphi\text{ and } \forall y^\subphi\in \spaceY,d_{\spaceY}(y,y^\subphi)\leq \epsilon,\\
0, \quad \text{otherwise,}
\end{array}\right.
\end{equation}
\end{tightdisplay}where $y^\subphi:=\{y\in\spaceY| L(y)\vDash \subphi\}$.

\par Let $\DFA=(Q,q_0,\Sigma_\DFA,\tau_\DFA,q_f)$ be a DFA with Boolean-formula-labeled transitions. For $\epsilon \geq 0$ we define robust indicator functions $\mathcal{L}_{\subphi}^{\epsilon}$ as in \eqref{eq:robust_indicator}. The $\epsilon$-robust satisfaction probabilities can be obtained directly as in Proposition~\ref{prop:spec_satisfaction} based on computing value functions iteratively using operators $\op_\subphi^\epsilon$ ($\mathcal{L}_\subphi$ in \eqref{eq:op_general} replaced by \eqref{eq:robust_indicator}).

\par
\noindent {\bfseries Section summary.}\hspace{1mm}In this section, we introduced $\delta$-lifted approximate simulation and $\epsilon$-robust indicator functions, implying that $(\epsilon,\delta)$-robust satisfaction probability can be computed. The next section shows how an a posteriori correction refines value functions back to certified lower bounds for the concrete continuous system. For ease of presentation, we assume $\epsilon=0$ for the next two sections. The extension to $\epsilon >0$ follows directly by replacing nominal indicator functions with robust ones, such that acceptance on the approximate model implies acceptance on the concrete system under $(\epsilon,\delta)$-similarity relation. This extension leaves all of our main results intact.

\section{Tensor trees, value iteration, and  a-posteriori correction}\label{sec:tensor_tree_ap}
In this section, we revisit the tree-structured tensor value-iteration method, as introduced in our earlier work \cite{wang2025unraveling}, for efficient computation of satisfaction probabilities of the abstract system $\hat{\mdpM}$. We will then show that these satisfaction probabilities can be corrected a-posteriori to compensate for approximation-induced robustness gaps. The resulting corrected values provide a provable lower bound on the satisfaction probabilities of the concrete system $\mdpM$.
\subsection{Tensor and tensor tree} \label{sec:subsec:cdc}
For $\spaceX=\prod_{i=1}^{N}\spaceX_i$, the value functions $\tensorV_{q,t}^{\boldsymbol{\pi}}:\spaceX  \rightarrow [0,1]$, can also be represented by an order-$N$ tensor, which is a multi-dimensional array in $ \mathbb{R}^{n_1\times n_2\times \ldots \times n_N}$. 
%
%
 We say that there exists a \textbf{Canonical Polyadic Decomposition }(CPD) representation of $\tensorV_{q,t}$ \cite{hitchcock1927expression}, if there exists arrays $v^{(i)}(z)\in \mathbb R^{n_i}$ s.t. 
	\begin{align}
	\tensorV_{q,t}= \sum_{z=1}^{|\mathcal{Z}|} v^{(1)}(z) \outerproduct v^{(2)}(z) \outerproduct \ldots \outerproduct v^{(N)}(z)
	\end{align}
	where $\outerproduct$ denotes the outer product  
	 and where the CPD rank of $	\tensorV_{q,t}$ is $|\mathcal{Z}|\in \mathbb{N}^+$ and is denoted as $\operatorname{rank}(	\tensorV_{q,t})$.
\ruohan{
Note that not every tensor admits a CP decomposition, and even when it does its CP rank is generally difficult to determine \cite{rozada2024tensor}. In our framework, the polyadic representation is a computational factorization of the joint abstraction instead of an additional assumption on the controller. This factorization arises naturally for dynamically decoupled components and can be extended to sparsely interconnected networks via compositional abstraction results \cite{schon2023verifying}. 
}
We are interested in doing value iterations based on low-rank CPD representations of our value functions. To this end, we introduce the notion of a tensor-tree value function. 
\begin{definition}[Tensor-tree value function $\mathcal{G}$ \cite{wang2025unraveling}] \label{def:r1tree}
\hspace{1mm}For a given DFA  $\DFA=(Q,q_0,\Sigma_\DFA,\tau_{\DFA},q_f)$, 	a tensor-tree value function 
	$\mathcal{G}=(\mathcal{Z},\mathcal{E},\qmapping, \valuemapping)$  has \begin{itemize}
			\item a set of vertices $\mathcal{Z}$ with elements $z\in \mathcal{Z}$; 
			\item a set of (labelled) edges $\mathcal{E}$ with elements $ (z,\subphi,z')\in\mathcal{E} $  that defines a \emph{rooted tree}.  
			\item a DFA-state mapping $\qmapping:\mathcal{Z}\rightarrow Q$ that maps a vertex $z\in\mathcal{Z}$ to a DFA state $q\in Q$.
			\label{def:expanding_tree}
			\item a vertex value mapping   
	$\valuemapping:\mathcal{Z}\rightarrow \mathbb{R}^{\prod |\spaceX_i|}$  that maps a vertex $z\in \mathcal{Z}$ to a \textbf{rank-1 tensor} 
	\begin{align*}\valuemapping(z)  = \bigotimes_{i=1}^N v^{(i)}(z)
		 .\end{align*}

	\end{itemize}
	For a given DFA mode $q$, the tensor-tree value function maps to a value function 
	\begin{align*}
		\tensorV_q = \sum_{z:\qmapping(z)=q} \valuemapping(z) \in  \mathbb{R}^{\prod |\spaceX^{(i)}|}, 
	\end{align*}
with 	$\tensorV_{q}=\boldsymbol{0}$ if $\{z | \qmapping(z)=q\}=\emptyset$.
\end{definition}

\begin{algorithm}[t]
	\caption{Tensor-tree value iteration $\mathcal{T}^{\pi}(\mathcal{G})$ \cite{wang2025unraveling}}\label{alg:sofie}
	\begin{algorithmic}[1]
		\Procedure{$\mathcal{T}^{\pi}$}{$\mathcal{G}$}
		\State $\mathcal{G}^+\leftarrow \operatorname{Grow}(\mathcal{G})$	\Comment{Grow tree}
		\For{  $(z,\subphi,z')\in\mathcal G^+.\mathcal{E}$}		\Comment{Update tree values}
		\State $\ruohan{\mathcal{G}^+.\valuemapping(z')=\op_{\subphi}^{\pi}(\mathcal{G}.\valuemapping(z))}$ 
		\EndFor
		
		\EndProcedure
	\end{algorithmic}  
\end{algorithm}

We show that the tensor-tree format is preserved under the recursive computation of $\tensorV_{q,t}$ in \eqref{eq:Viterations} if the strategy of the agents only depends on their own state and the mode $q$, that is if the policy   $\boldsymbol{\pi}=\{\pi[t]|t=0,1,\ldots,T-1\}$ has  elements $\pi[t]\in\Pi_{\mathbf{M}}$, denoted as $\boldsymbol{\pi}\in \Pi_{\mathbf{M}}$, and defined as
\begin{equation}\label{eq:PiM}
\Pi_{\mathbf{M}}:= \{ \pi =\operatorname{col}(\pi^{(1)},...,\pi^{(N)}) | \pi^{(i)}:\spaceX_i\times Q\rightarrow \spaceU_i \}.
\end{equation}

 First note that the initial value function $\tensorV_{q,0}$ can be trivially represented by   
 $\mathcal{G}_0=\{\mathcal{Z}_0,\mathcal{E}_0,{\valuemapping}_0,{\qmapping}_0\}$, with $\mathcal{Z}_0=\{1\}$, $\mathcal{E}_0=\emptyset$, ${\valuemapping}_0(1)=\bigotimes_{i=1}^N \indicator$, and $\qmapping(1)=q_f$. 
 Based on Algorithm \ref{alg:grow}, we can grow the number of leafs 
$\mathcal{G}_0^+= \operatorname{Grow}(\mathcal G_0)$ after which we  can compute the updated vertex values based on the operator 
$ \op_{\subphi}^{\pi_q} $  in  eq. \eqref{eq:op_general}.  
\begin{algorithm}[tp]
\caption{Growing tensor tree $\operatorname{Grow}(\mathcal{G})$ \cite{wang2025unraveling}}\label{alg:grow}
\begin{algorithmic}[1]
	\Procedure{$\operatorname{Grow}$}{$\mathcal{G}$}
	\Comment{Grow tree}
	\For{ $z \in \mathcal{G}.\textit{leaves}$ } 
	\For{  $\{(q,\subphi,q')\in\tau_\DFA|q'=\mathcal{L}_Q(z)\}$}
	\State $\bar{z} \leftarrow$ CreateNewVertex
	\State $\mathcal G.\mathcal{Z}\leftarrow\mathcal{Z}\cup \{\bar z\}$ 
	\State $\mathcal G.\mathcal{E} \leftarrow \mathcal{E}\cup \{(z,\subphi,\bar{z})\}$
	\State $ \mathcal G.\mathcal{L}_Q(\bar{z}) \leftarrow  q$
	\EndFor
	\EndFor 
%
	
	\EndProcedure
\end{algorithmic}  
\end{algorithm}
 For a policy sequence $\boldsymbol{\pi}=\{\pi[0],\ldots,\pi[T-1]\}$, this yields the tensor-tree value iteration defined in Alg.\ref{alg:sofie} with
\begin{align}
    \mathcal{G}_{t+1}=\optree^{\pi[T-1-t]}(\mathcal{G}_t).
    \label{eq:tree_expansion_iteration}
\end{align}
As in \cite{wang2025unraveling}, we can show that the tensor-tree value iterations preserve vertices with rank-1 tensor values with whose value the value function in can be recovered. 
\begin{proposition}
	[Tensor-tree value iterations \cite{wang2025unraveling}] \label{theorem:tree_to_value_function} 
\hspace{1mm}For a horizon $T$, a Markov policy 
	$\boldsymbol{\pi}=\{\pi[0],\ldots,$ $\pi[T-1]\}\in\Pi_{\mathbf{M}}$ as in \eqref{eq:PiM}
	and a tree $\mathcal{G}_t=\{\mathcal{Z}_t,\mathcal{E}_t,{\qmapping}_t,{\valuemapping}_t\}$ computed with tensor-tree value iterations \eqref{eq:tree_expansion_iteration}, we have that 
	\begin{itemize}
		\item $\mathcal{G}_t$ is a rank-1 tensor tree as defined in Def. \ref{def:r1tree}, that is, $$\operatorname{rank}({\valuemapping}_t(z)) =1 \quad \forall z\in\mathcal{Z} $$
		\item  $\mathcal{G}_t$ represents the  value functions $\tensorV_{q,t}^{\boldsymbol{\pi}} \quad \forall q\in Q$, computed based on \eqref{eq:Viterations}, that is,
		\begin{align}
			\tensorV_{q,t}^{\boldsymbol{\pi}}(x)=\sum_ {z:\qmapping(z)=q}v_t(z)(x) \quad \forall x\in \spaceX, \forall q\in Q.  
			\label{eq:compute_tensorv_based_on_tree}
		\end{align}
	\end{itemize}

\end{proposition}
\ruohan{Compared to monolithic dynamic programming, tensor-tree avoids storing value functions over the full joint state space $|Q|\prod_{i=1}^m |\spaceX_i|$ and reduces the memory usage from multiplicative  $O(|Q|\prod_{i=1}^m |\spaceX_i|)$ to additive $O(\text{Rank}\sum_{i=1}^{m} |\spaceX_i|)$, where $\text{Rank}:=|\{z | \mathcal{L}_Q(z)=\tau_\DFA(q_0,x_0)\}|$ is a DFA-dependent factor and can be reduced by applying pruning Alg.~2 in \citep{wang2025unraveling}.
}
\par
\noindent {\bfseries Policy optimization.}\hspace{1mm}As in \cite{wang2025unraveling}, for each mode $q$ and each subsystem $i\in \{1,\ldots,N\}$, we can  develop efficient heuristic optimization of the policy with
\vspace{-2mm}
\begin{tightdisplay}
 \begin{equation} 
 	\begin{aligned}
 	\pi_q^{ (i)\ast}(x_i,u_i)	\in \arg\max_{\pi_q^{(i)}} \sum_{e\in\mathcal E_{q}} \op_{\subphi^{i}}^{	\pi_q^{(i)} }(v^{(i)})(z) c_{e}^{i}
 	\end{aligned} \label{eq:pol_base}
 \end{equation}
\end{tightdisplay}where $\mathcal E_{q}:=\{e =(z,\alpha,z')\in \mathcal E| \qmapping(z') = q \}$ and $c^{i}_{e}:= \prod_{j}^{\{1,\ldots,N\} \setminus \{i\}} \|\op_{\subphi^{j}}^{	\pi_q^{(j)} }(v^{(j)})(z) \|_1.$
\color{black}
\subsection{A-posteriori corrected satisfaction probability } \label{subsec:apos}
Given a multi-agent gMDP $\mdpM$ as in \eqref{eq:cart_gmdp}, we require that each agent $\mdpM^i$  has an abstract agent $\hat{\mdpM}^i$ such that for all $i\in \{1,\ldots,N\}$, $\hat{\mdpM}^i \preceq_{0}^{\delta^i} \mdpM^i$. Together, these abstract gMDPs define the abstract multi-agent gMDP $\hat{\mdpM}$ which is $(\boldsymbol{\delta},0)$-simulated by $\mdpM$ with $\boldsymbol{\delta}:=1-\prod_{i=1}^{N}(1-\delta^i)$. 
\par 
Leveraging the tensor-tree value iterations in Algorithm \ref{alg:sofie}, we can efficiently use the tensor tree to compute value functions $\tensorV_{q,T}^{\boldsymbol{\pi}}(\hat x)$. 
 In the rest of this section, we use the approximate simulation relation between $\mathbf M$ and $\hat{\mathbf{M}}$ to relate the abstract computations back to satisfaction probabilities of the concrete multi-agent system $\mdpM$. Under the adopted similarity quantification, probability losses accumulate up to the acceptance event. Its effect can be quantified with the \textit{first hitting time} \cite{haesaert2020robust}, defined as follows.
For a given execution $\boldsymbol{\hat{x}}$, the first hitting time of a set $\{q_f\}$ is a random variable conditioned on the initial state $\hat{x}_0$ and defined as
\begin{align} \label{eq:fht}
    H_{\{q_f\}} (\hat{x}_0)\!:=\!\bigl \{ \inf\{t\in \mathbb{N} \union\! \{\infty\}\! :\! q_t\! \in\! \{q_f\}\} \!-\!1| \hat{x}_0  \bigr\},
\end{align}
where $q_{k+1}:=\tau_\DFA(q_k,L(\ruohan{g}(\hat{x}_k)))$ for $k\in \mathbb{N}$, with $q_0:=q_0$. The support of first hitting time is $\mathbb{N} \union \{\infty\}$. It can be infinity if the execution does not hit the set $\{q_f\}$. $ H_{\{q_f\}}(\hat{x}_0)$ is $0$ with probability $\mathbb{P}(H_{\{q_f\}}(\hat{x}_0)\!=\!0)\!=\!1$ if $\tau_\DFA(q_0,L(\ruohan{g}(\hat{x}_0)))=q_f$.
%
%
%

\begin{proposition}[$\mathbf{C}_{\boldsymbol{\pi}}$ with satisfaction probability]\label{prop:fht}
\hspace{1mm}For multi-agent gMDP $\M$ and its abstraction $\hat\M$ with $\hat\M\preceq^{\delta}_0\M$, we can \emph{construct} a control strategy $\mathbf{C}_{\boldsymbol{\pi}}$ such that the controlled system $\mdpM \times \mathbf{C}_{\boldsymbol{\pi}}$ satisfies $\phi$ with lower bound
   \begin{equation}\notag
      \begin{aligned}
          \label{eq:fht_inequa}
          &	\ruohan{\mathbb{P}^{x_0}_{\!\mdpM \times \mathbf{C}_{\boldsymbol{\pi}}}\!( \word_{[0,T]} \models \phi ) } \!\geq\! \tensorV_{\bar{q}_0,T}^{\boldsymbol{\pi}}(\hat x_0)   \!-\!\boldsymbol{\delta} \sum_{h=1}^{T} \mathbb{P}(H_{\{q_f\}}(\hat{x}_0) \geq h) ,
      \end{aligned}
      \end{equation}
      for any $T \geq 1$ with $\bar q_0= \tau_{\mathcal A}(q_0,L(\ruohan{g}(\hat x_0)))$. 
\end{proposition}
\begin{proof}
The proof follows the same reasoning as in the proof of Thm. 3 \cite{haesaert2020robust} and is omitted here due to space limitation. 
\end{proof}
In general, $\mathbb{P}(H_{\{q_f\}}(\hat{x}_0) \geq h)$ can be very difficult to compute. However, given the information encapsulated in the tensor tree we can compute a bound on it as follows. 
\begin{theorem}[$\mathbf{C}_{\boldsymbol{\pi}}$ with a-posteriori corrected satisfaction probability]
\label{theorem:branch_length_lossProb}
\hspace{1mm}For multi-agent gMDP $\M$ and its abstraction $\hat\M$ with $\hat\M\preceq^{\delta}_0\M$, we can \emph{construct} a control strategy $\mathbf{C}_{\boldsymbol{\pi}}$ such that the controlled system $\mdpM \times \mathbf{C}_{\boldsymbol{\pi}}$ satisfies $\phi$ with lower bound
          \begin{equation}
         \begin{aligned}
           &\!\!\!\!	\ruohan{\mathbb{P}^{x_0}_{\!\mdpM \times \mathbf{C}_{\boldsymbol{\pi}}}}\!( \word_{[0,T]} \!\!\models\!\! \phi ) \!\geq\! \tensorV_{\bar{q}_0,T}^{\boldsymbol{\pi}}(\hat x_0)   \!-\!\boldsymbol\delta(T\!-\!\!\! \sum_{h=1}^{T}\!\sum_{i=0}^{h-1}\!\!\!\!\!\sum_{\ \ z\in\mathcal{D}_i^{\bar q_0} }\!\!\!\!v(z))\!\!\!\!\!\!\!
         \end{aligned}
     \end{equation}
     \noindent for any $T \geq 1$ with $\bar q_0= \tau_{\mathcal A}(q_0,L(\ruohan{g}(\hat x_0)))$  where $v(z)$ is the value of the vertex that belongs to $\mathcal D_i^{\bar q_0}$.  $\mathcal{D}_i^{\bar q_0}$ is the set containing $\bar q_0$-labeled vertices, each of which has exactly $i$ predecessors, defined as
        \begin{align} \label{eq:same_predecessor}
   \!\!\! \mathcal{D}_i^{\bar q_0} \!= \!\{z| \textmd{$z$ has  $i$ predecessors and $\mathcal{L}_Q(z)\!=\!{\bar q_0}$}  \}.\!\!\!\!
    \end{align}

\end{theorem}
\vspace{-2mm}
\begin{proof}
    It is trivial that
      $\mathbb{P}(H_{\hat{\spaceX} \times \{q_f\}}(\hat{x},q) \!\geq \! h) \!= \!1\!- \!\mathbb{P}(H_{\hat{\spaceX} \times \{q_f\}}(\hat{x},q) \!<\! h). $ 
    Based on Proposition~\ref{prop:fht}, we get
   \begin{align}
        \tensorV^{\boldsymbol{\pi}}_{q,T}(\hat{x},x) \geq \tensorV^{\boldsymbol{\pi}}_{q,T}(\hat{x})-\boldsymbol{\delta} \sum_{h=1}^{T} \biggl(1-\mathbb{P}(H_{\hat{\spaceX} \times \{q_f\}}(\hat{x},q) < h) \biggr), 
   \nonumber\end{align}
    where $\mathbb{P}(H_{\hat{\spaceX} \times \{q_f\}}(\hat{x},q) < h) $ denotes the probability the controlled system product reaches the accepted state within $h-1$ control steps. Each vertex $v(z)$ with $h-1$ predecessors defines the probability of a unique accepted path of length $h-1$, therefore, such probability can be lower bounded by the vertices' values, i.e.,
    \begin{tightdisplay}
                    $\mathbb{P}(H_{\hat{\spaceX}\times \{q_f\}}(\hat{x},q) < h) \geq \sum_{i=0}^{h-1}\sum_{z\in\mathcal{D}_i^q}v(z),$ 
    \end{tightdisplay}where $\mathcal{L}_{Q_{h-1}}^{-1} (q):Q\rightarrow \mathcal{Z} $ gives the vertices set in $\mathcal{G}_{h-1}$ which is labeled $q$, and $\mathcal{D}_i^q$ is the set containing $q$-labeled vertices, each of which has exactly $i$ predecessors, defined as in \eqref{eq:same_predecessor}.
\end{proof}
\noindent {\bfseries A-posteriori corrected policy optimization.}\hspace{1mm}For each mode $q$ and each agent $i$, we extend the policy optimization in \eqref{eq:pol_base} for $\pi_q^i\in \Pi_{\mathbf{M}}$ to account for the a-posteriori correction by multiplying a weighted term as follows.
\vspace{-2mm}
\begin{tightdisplay}
\begin{equation}
	\begin{aligned} \label{eq:pol_max_apos}
	\pi_q^{(i)\ast}(x_{i},u_{i})	\in \arg\max_{\pi_q^{(i)}} \sum_{e\in\mathcal E_{q}} \zeta_{z'} \op_{\subphi^{i}}^{	\pi_q^{(i)}}(v^{(i)})(z)   c_{e}^{i}
	\end{aligned}
\end{equation}
\end{tightdisplay}where $\mathcal E_{q}:=\{e =(z,\alpha,z')\in \mathcal E| \qmapping(z') = q \}$, $\zeta_{z'}:=1+\boldsymbol{\delta}(T-{\small \text{Pred}}(z'))$ with ${\small \text{Pred}}(z')$ the number of predecessors of $z'$, and 
$c^{i}_{e}:= \prod_{j}^{\{1,\ldots,N\} \setminus \{i\}} \|\op_{\subphi^{j}}^{	\pi_q^{(j)} }(v^{(j)})(z) \|_1 $.

\color{black}
\section{Robust tensor tree} \label{sec:new_robust_tensor_tree}
This section presents a robust version of \eqref{eq:Viterations} by proposing a robust operator via which we embed probability loss quantification into value iteration, following the same idea as $\delta$-robust mapping in \cite{haesaert2020robust}. Unlike the a-posteriori correction as introduced in the previous section, this online robust correction accounts for probability loss during iteration rather than until acceptance, such that guaranteed lower bounds are guaranteed throughout recursive computation.

\begin{algorithm}[t]
	\caption{Robust tensor-tree value iteration $\mathcal{T}^{\pi}(\mathcal{G})$ }\label{alg:robusttree}
	\begin{algorithmic}[1]
		\Procedure{$\mathcal{T}^{\pi}$}{$\mathcal{G}$}
		\State $\mathcal{G}^+\leftarrow \operatorname{Grow}(\mathcal{G})$	\Comment{Grow tree}
		\For{  $(z,\subphi,z')\in\mathcal G^+.\mathcal{E}$}		\Comment{Update tree values}
		\State $\ruohan{\mathcal{G}^+.\valuemapping^{(i)}(z')=\op_{\delta^{i},\subphi^{i}}^{\pi^{i}}(\mathcal{G}.v^{(i)})(z)}$ $\forall i\in \{1,\ldots, N\}$
		\EndFor
		
		\EndProcedure
	\end{algorithmic}  
\end{algorithm}

We define the $\delta^i$-robust DFA-informed operator $\op^{\delta^i}$ as 
\begin{equation}
	\begin{aligned}
		&\op_{\delta^{i},\subphi^{i}}^{\pi}(v^{(i)})(x_{i})
:=\trunc \biggl( \expectation_{x'_{i}}^{\pi_q^i}[\mathcal{L}_{\subphi^{i}}(x_{i}') v^{(i)}(x'_{i})]-\delta^{i} \biggr).
	\end{aligned}
	\label{eq:subagent_robust_operator}
\end{equation}
\vspace{-0.2cm}

Unlike in the previous subsection, we present the proposed operator explicitly for each $\mdpM^i$ because $\delta^i$ is $\mdpM^{i}$-specific; as reusing the generic form \eqref{eq:op_general} would obscure this dependence. 
We perform robust dynamic programming on $\hat{\mdpM}$ by leveraging the robust tensor-tree value iteration in Algorithm~\ref{alg:robusttree}. We use the approximate relation between $\mdpM$ and $\hat{\mdpM}$ to relate the computed robust value functions back to satisfaction probabilities of the controlled multi-agent system $\mdpM\times \mathbf{C}_{\boldsymbol{\pi}}$. The robust tensor tree provides a direct certified lower bound as follows. 

\begin{theorem}[$\mathbf{C}_{\boldsymbol{\pi}}$ with robust-tree corrected satisfaction probability] \label{theorem:satprob_lb_robusttree}
\hspace{1mm}For multi-agent gMDP $\mdpM$ and its abstraction $\hat{M}$ with $\hat{\mdpM}\preceq_{0}^{\delta} \mdpM$, we can construct a control strategy $\mathbf{C}_{\boldsymbol{\pi}}$ such that the controlled system $\mdpM \times \mathbf{C}_{\boldsymbol{\pi}}$ satisfies $\phi$ with lower bound
\begin{align} \label{eq:theorem:satprob_lb_robusttree}
    \ruohan{\mathbb{P}^{x_0}_{\mdpM \times \mathbf{C}_{\boldsymbol{\pi}}}(\word_{[0,T]} \vDash \phi )} \geq  \sum_{n\in {\qmapping}_T^{-1}(\bar{q}_0)}v_T^{\delta}(z)
\end{align}
\noindent for any $T\geq 1$ with $\bar{q}_0:=\tau_\DFA(q_0,L(\ruohan{g}(\hat{x_0})))$, where ${\qmapping}_T^{-1}(\bar{q}_0):Q\rightarrow \mathcal{Z}$ gives the vertice set in 
	$\mathcal{G}_{T}^{\delta}$ which is labeled by $\bar{q}_0$. $\mathcal{G}_T^{\delta}$ is robustly expanded as in Algorithm~\ref{alg:robusttree}.  
\end{theorem}
\vspace{-3mm}
\begin{proof}
    The proof of Theorem~\ref{theorem:satprob_lb_robusttree} follows along the same lines as the proof of Theorem~2 in \cite{haesaert2020robust} with $\bigotimes_{i=1}^N \op_{\delta^i,\subphi^i}^{\pi_q^i}$ instead of $\op$. For a complete proof, we refer to the extended version of this paper \cite{wang2025robusttensor}.
\end{proof}

\vspace{-0.2cm}

\noindent {\bfseries Robust-tree policy optimization.}\hspace{1mm}To account for the robust-tree correction, for each mode $q$ and each agent $i$, we optimize policy $\pi_q^i\in \Pi_{\mdpM}$ as\vspace{-2mm}
\begin{tightdisplay}
\begin{equation}
	\begin{aligned} \label{eq:maxpol_rt}
	\pi_q^{(i)\ast}(x_i,u_i)	\in \arg\max_{\pi_q^{(i)}} \sum_{e\in\mathcal E_{q}} \op_{\subphi^{i}}^{	\pi_q^{(i)},\delta^i }(v^{(i)})(z) c_{e}^{i}
	\end{aligned}
\end{equation}
\end{tightdisplay}where $\mathcal E_{q}:=\{e =(z,\alpha,z')\in \mathcal E| \qmapping(z') = q \}$ and $c^{i}_{e}:= \prod_{j}^{\{1,\ldots,N\} \setminus \{i\}} \|\op_{\subphi^{j}}^{	\pi_q^{(j)},\delta^j }(v^{(j)})(z) \|_1$. Compared to policy optimization in \eqref{eq:pol_max_apos}, this correction uniformly subtracts $\delta^i$ for all nodes and all states, and is therefore more conservative.
\vspace{-0.2cm}

\color{black}
\section{Benchmarking}\label{sec:case_study}
\vspace{-1em}
All simulations were conducted in Python on a MacBook Air with Apple M4 chip, and 16 GB RAM.

\begin{table}[!h]
\centering
\setlength{\tabcolsep}{3pt}
\caption{\begin{small}Per-iteration required analytical memory (bytes) of the value function. $\mathrm{Rank}_t$ denotes the number of rank-1 terms in the tensor-tree representation at iteration step $t$, $\mathrm{Rank}_t\in [1,476]$ in discussed benchmarks.\end{small}}

\begin{tabular}{p{1.5cm} p{0.3cm} p{0.8cm} c c}
\hline
\begin{footnotesize}Benchmark\end{footnotesize} & \begin{footnotesize}$N$\end{footnotesize} & \begin{small}$|\spaceX_i|$\end{small} & \begin{footnotesize}Monolithic DP [bytes]\end{footnotesize} & \begin{footnotesize}Tensor-tree [bytes]\end{footnotesize} \\
\hline
\ref{sec:cs_ra} & 2  & 1000 & $2.4\times 10^7$     & $\mathrm{Rank}_t \times 1.6\times 10^{4}$ \\
\ref{sec:cs_IV} & 20 & 100  & $2.4\times 10^{41}$  & $\mathrm{Rank}_t \times 1.6\times 10^{4}$ \\
\ref{sec:cs_pd} & 8  & 1200 & $1.02\times 10^{26}$ & $\mathrm{Rank}_t \times 7.7\times 10^{4}$ \\
\hline
\end{tabular}
\label{tab:mem_comp}

\vspace{-0.2cm}

\end{table}
\subsection{Baseline Validation}\label{sec:cs_ra}
\begin{figure}[!t]
  \centering
  \includegraphics[width=1\columnwidth,trim=0 34pt 0 0,clip]{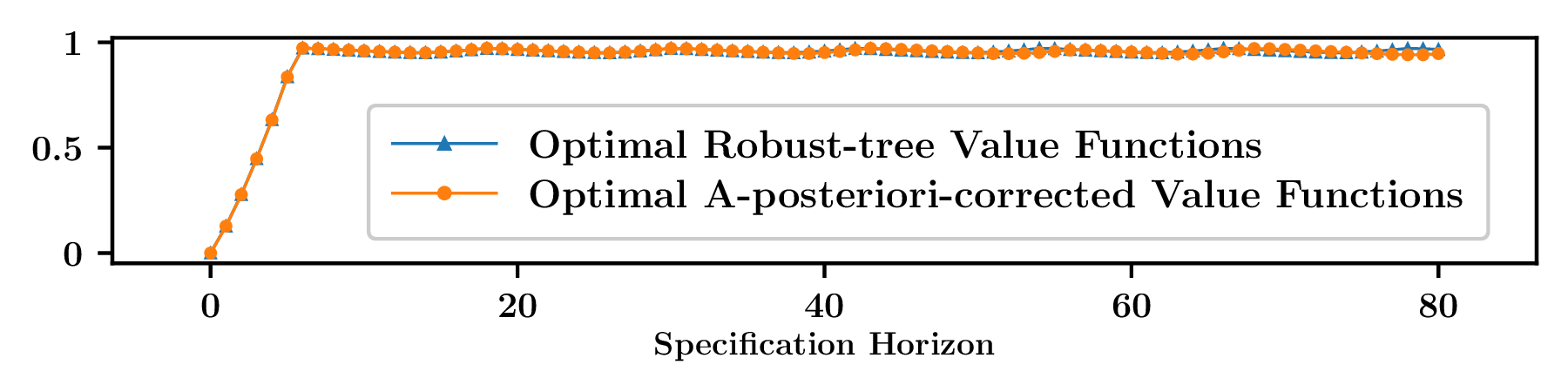}
  \caption{\begin{small}\ruohan{Lower bounds on the satisfaction probability of a reach-avoid specification as a function of the specification horizon. Orange: a-posteriori corrected (Thm.~\ref{theorem:branch_length_lossProb}, $\Pi_{\mdpM}^\ast$ optimized as in \eqref{eq:pol_max_apos}); blue: robust-tree corrected (Thm.~\ref{theorem:satprob_lb_robusttree}, $\Pi_{\mdpM}^\ast$ optimized as in \eqref{eq:maxpol_rt}).}\end{small}}
\label{fig:RA_2D_2CURVES}
 \vspace{-0.1cm}
\end{figure}

\vspace{-0.8em}
We first consider a moderate-scale two-agent reach-avoid benchmark to validate the proposed framework and to compare the two computational approaches to account for the robust satisfaction probabilities. We consider each agent's dynamics as Eq.~(17) and output $y_i=x_i$ and a reach-avoid specification as in Ex.~1 in \cite{wang2025unraveling}.
We construct a finite-state abstraction for each agent $\hat{\mdpM}^1,\hat{\mdpM}^2$ by gridding the state space $\spaceX_i$ with $1000$ grid cells and the input space $\spaceU_i$ with $10$ grid cells. With a chosen interface function $u_i=\hat{u}_i$ and relation $\relation_i$ with $\epsilon^i=0.1$, we obtain the approximate simulation relation $\hat{\mdpM}^i\preceq_{0.1}^{0.002} \mdpM^i$ using \textbf{SySCoRe} tool \cite{van2023syscore}. We compute the average of probabilities for system states initialized as $x_0\in[-20,0]\times[-5,-2.5]$, visualized in Fig.~\ref{fig:RA_2D_2CURVES}. \ruohan{The lower bounds on the satisfaction probabilities obtained by robust-tree and aposteriori-correcion both converge for this reach-avoid specification.}


\begin{figure}[htbp]
  \centering
  \includegraphics[width=1\columnwidth,trim=0 38pt 0 14pt,clip]{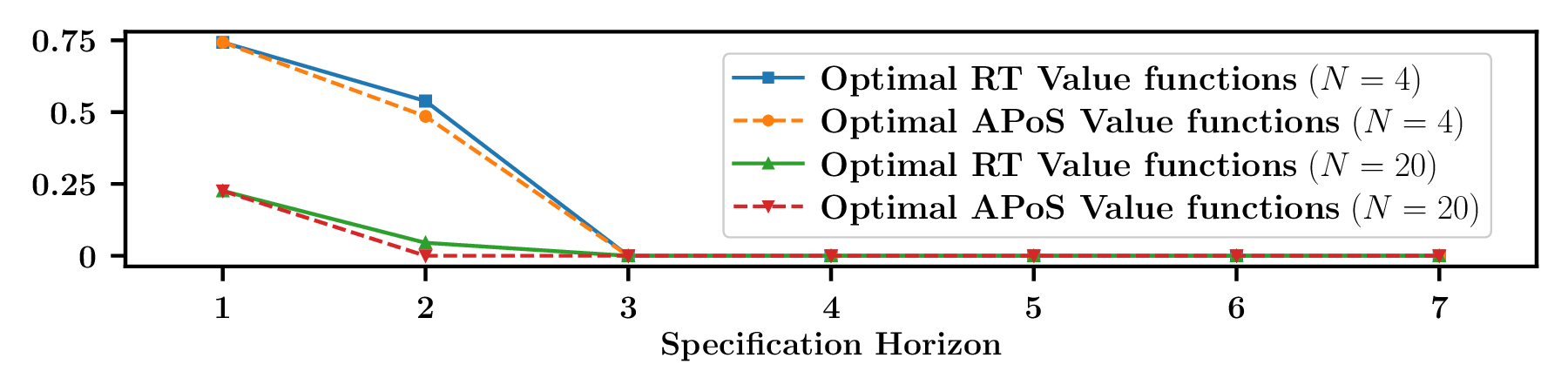}
  \caption{\begin{small}Lower bounds on the satisfaction probability of an invariance specification on a system as a function of the specification horizon $T$. Orange, red: a-posteriori (\textbf{APoS}) corrected (Thm.~\ref{theorem:branch_length_lossProb}, $\Pi_{\mdpM}^\ast$ optimized as in \eqref{eq:pol_max_apos}) for $4$-agent system and $20$-agent system; blue, green: robust-tree (\textbf{RT}) corrected  (Thm.~\ref{theorem:satprob_lb_robusttree}, $\Pi_{\mdpM}^\ast$ optimized as in \eqref{eq:maxpol_rt}) for $4$-agent system and $20$-agent system.\end{small}}
\label{fig:sca_safe_T20}
\vspace{-0.2em}
\end{figure}

\subsection{System Size Scalability}\label{sec:cs_IV}
\vspace{-0.8em}
We then consider a large-scale 20-agent benchmark to highlight our approach's scalability in terms of system size, particularly in regimes where the size renders monolithic dynamic programming impractical as justified in \ruohan{Table~\ref{tab:mem_comp}}.
We consider each agent's dynamics as in subsection~\ref{sec:cs_ra}
with $N=20$, $w_{i}\sim \mathcal {N}(0,I)$, $\spaceX_{i}:[-10,10]$, $\mathbb{U}_{i}:[-2,2]$. Consider an invariance specification $\phi:=\bigwedge_{t=0}^{T} \bigcirc(\bigwedge_{i=1}^{N=20}   p_1^{i})$ with  $p_1^i$ is true iff $x_i\in [-5,5]$. We construct a finite-state abstraction for each agent by gridding the state space $\spaceX_i$ with 100 grid cells and the input space $\spaceU_i$ with 10 grid cells. We choose interface function $u_i=\hat{u}_i$ and relation $\relation_i$ with $\epsilon^i=0.1$. We obtain $\hat{\mdpM}^i\preceq_{0.1}^{0.0717} \mdpM^i$ using \textbf{SySCoRe} tool \cite{van2023syscore}. We present the average of (5 sampled initial states) of the lower bounds of satisfaction probability computed via a-posteriori as in Thm.~\ref{theorem:branch_length_lossProb} and robust tree as in Thm.~\ref{theorem:satprob_lb_robusttree} in Fig.~\ref{fig:sca_safe_T20} for a sweep of specification horizon, i.e., $T=1,\ldots,7$. It is expected that as the number of agents increases, the probability that all agents remain in the invariant set over a fixed horizon decreases, highlighting the difficulty in satisfying an invariance specification as number of agents grows. However, robust computations do not rely on all agents staying in the simulation relation are therefore of interest for future work.



\subsection{Cyclic DFA Scalability}\label{sec:cs_pd}
\vspace{-0.8em}
We consider a 4-robot-8-dimensional package delivery task with a cyclic DFA specification as in Fig.~\ref{fig:PD_DFA} to demonstrate that the proposed method can handle complex temporal logic specifications beyond simple ones. We consider each dimension of each robot's dynamics as in subsection~\ref{sec:cs_ra}. For the considered specification, $\subphi_1,\subphi_2,\subphi_3,\subphi_4,\subphi_5$ are composed of atomic propositions from Robot A, B, C, and D. $\subphi_6,\subphi_{10}$ are composed of atomic propositions from Robot A, $\subphi_7,\subphi_{10}$ Robot B, $\subphi_8,\subphi_{10}$ Robot C, and $\subphi_9,\subphi_{10}$ Robot D, respectively. We visualize the output regions for which these subformulas are true in Fig.~\ref{fig:subformula}. We construct a finite-state abstraction for each dimension (of each robot) by gridding the state space $\spaceX_i$ with $1200$ grid cells and the input space $\spaceU_i$ with $10$ grid cells. We choose interface function $u_i=\hat{u}_i$ and relation $\relation_i$ with $\epsilon^i=0.1$. We obtain $\hat{\mdpM}^i\preceq_{0.1}^{0.0004} \mdpM^i$ using \textbf{SySCoRe} tool \cite{van2023syscore}. Out of $4.2998\times 10^{24}$ possible (and computed) combinations of initial states of 4 robots, we present for $3$ initial conditions in Table~\ref{tab:8D_prob} the lower bounds on the a-posteriori corrected satisfaction probabilities.

\begin{figure}[!t]
\centering
\resizebox{0.62\columnwidth}{!}{%
\begin{tikzpicture}[
    ->,
    >=stealth,
    shorten >=1pt,
    semithick,
    node distance=2.8cm,
    on grid,
    auto
]
    \node[state] (q0) {$q_0$};
    \node[state, right=of q0] (q1) {$q_1$};
    \node[state, accepting, right=of q1] (qf) {$q_f$};

    \draw[->] ($(q0.west)+(-1.2,0)$) -- (q0.west);

    \draw[->] (q0) -- node[above] {$\alpha_1$} (q1);
    \draw[->] (q1) -- node[above] {$\alpha_{10}$} (qf);
    \draw[->] (q0) edge[loop above] node {$\alpha_2$} ();

    \draw[->] (q1) edge[loop above] node {$\alpha_3 \mid \alpha_4 \mid \alpha_5$} ();

    \draw[->]
    (q1) to[out=235, in=305, looseness=0.9]
    node[midway, below=2pt] {$\alpha_6 \mid \alpha_7 \mid \alpha_8 \mid \alpha_9$}
    (q0);
\end{tikzpicture}
}
\caption{\begin{small}\ruohan{DFA of 4-robot package delivery specification.}\end{small}}
\label{fig:PD_DFA}
\vspace{-0.1cm}
\end{figure}

\begin{figure}[htp]
    \centering
    \includegraphics[width=1\columnwidth,trim=0 60pt 0 10pt,clip]{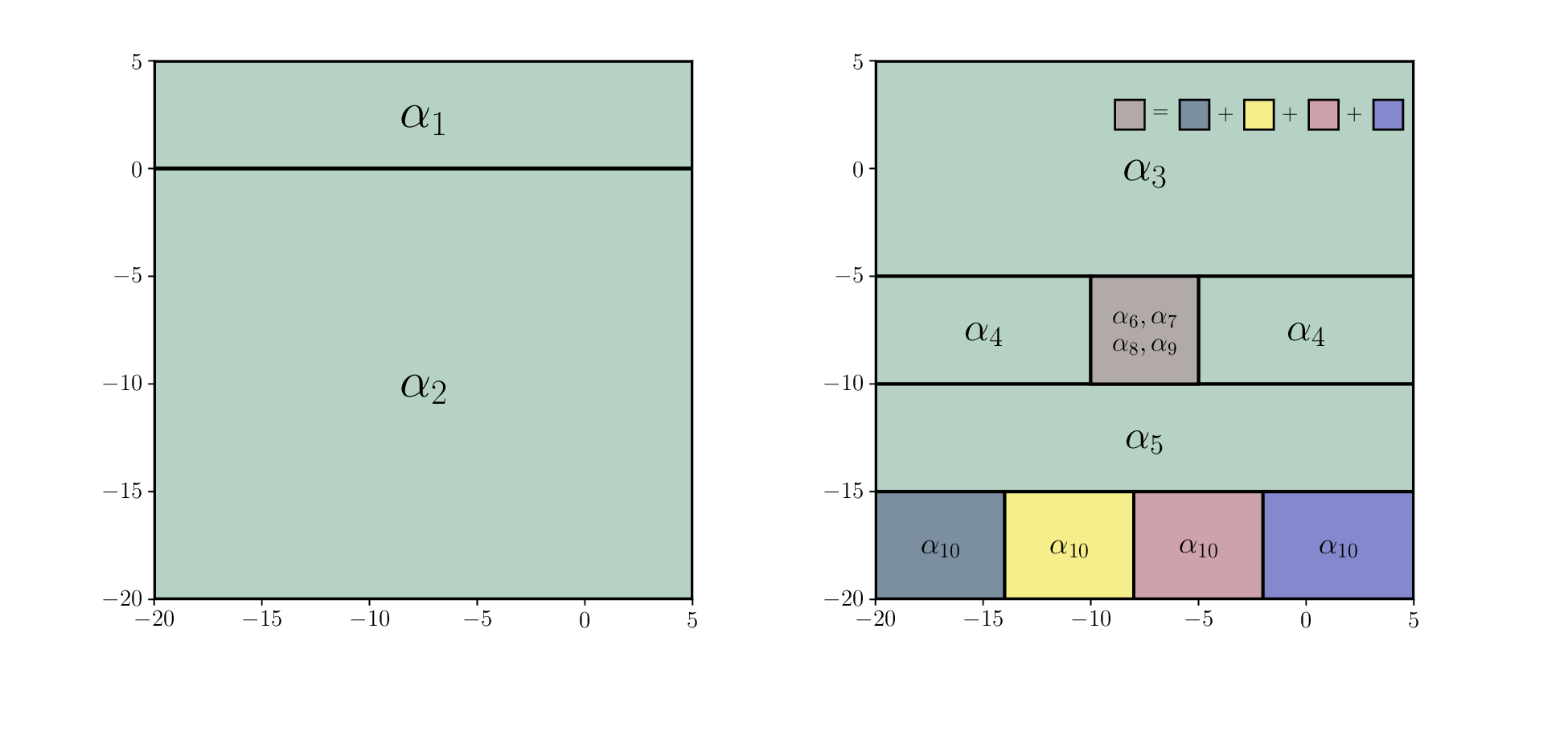}
    \caption{\begin{small}Subformula of DFA in Fig.~\ref{fig:PD_DFA} on output space of robots. \end{small}}
    \label{fig:subformula}
\end{figure}

\begin{table}[htp]
\centering
\setlength{\tabcolsep}{3pt}
\caption{\begin{small}\ruohan{A posteriori-corrected lower bounds on satisfaction probabilities for $3$ sets of robots' initial positions, $T=20$.}\end{small}}
{\footnotesize
\begin{tabular}{c | c c c }
\hline
$\mathbb{P}$ & $0.616$ & $0.615$ & $0.162$  \\
\hline
$x_{A,0}$ & $(1.4125,\,1.2875)$ & $(\text{-}17.8375,\,3.2875)$ & $(\text{-}18.0875,\,\text{-}7.2875)$ \\

$x_{B,0}$ & $(3.3625,\,1.2875)$ & $(\text{-}16.8625,\,3.2875)$ & $(\text{-}16.8625,\,\text{-}7.2875)$  \\

$x_{C,0}$ & $(\text{-}17.8375,\,1.2875)$ & $(1.4125,\,3.2875)$ & $(\text{-}13.5375,\,\text{-}7.2875)$  \\

$x_{D,0}$ & $(\text{-}16.8625,\,1.2875)$ & $(3.3625,\,3.2875)$ & $(\text{-}11.8625,\,\text{-}7.2875)$  \\
\hline
\end{tabular}

}
\label{tab:8D_prob}
\end{table}

\color{black}
\section{Conclusions}\label{sec:conclusions}
\vspace{-2mm}
This paper presents two correct-by-design control synthesis methods for continuous-state stochastic multi-agent systems that leverage tensor-structured value functions for scalable computation. Abstraction error is quantified via $(\epsilon,\delta)$ simulation relation and two ways of correction, i.e., a-posteriori and robust-tree propagation. We validate our results on linear systems. Future work includes specification that requires interactions among agents.
\vspace{-0.2cm}




\bibliography{references}

@inproceedings{rozada2024tensor,
  title={Tensor low-rank approximation of finite-horizon value functions},
  author={Rozada, Sergio and Marques, Antonio G},
  booktitle={ICASSP},
  pages={5975--5979},
  year={2024},
  organization={IEEE}
}

@inproceedings{schon2023verifying,
  title={Verifying the unknown: Correct-by-design control synthesis for networks of stochastic uncertain systems},
  author={Sch{\"o}n, Oliver and van Huijgevoort, Birgit and Haesaert, Sofie and Soudjani, Sadegh},
  booktitle={62nd Conf. Decis. Control},
  pages={7035--7042},
  year={2023},
  organization={IEEE}
}

@book{baier2008principles,
  title={Principles of model checking},
  author={Baier, Christel and Katoen, Joost-Pieter},
  year={2008},
  publisher={MIT press}
}

@article{hitchcock1927expression,
  title={The expression of a tensor or a polyadic as a sum of products},
  author={Hitchcock, Frank L},
  journal={J. Math. Phys.},
  volume={6},
  number={1-4},
  pages={164--189},
  year={1927},
  publisher={Wiley Online Library}
}

@article{haesaert2020robust,
  title={Robust dynamic programming for temporal logic control of stochastic systems},
  author={Haesaert, Sofie and Soudjani, Sadegh},
  journal={IEEE Trans. Autom. Control},
  volume={66},
  number={6},
  pages={2496--2511},
  year={2020},
  publisher={IEEE}
}

@article{liu2021symbolic,
  title={Symbolic models for infinite networks of control systems: A compositional approach},
  author={Liu, Siyuan and Noroozi, Navid and Zamani, Majid},
  journal={Nonlinear Anal. Hybrid Syst.},
  volume={43},
  pages={101097},
  year={2021},
  publisher={Elsevier}
}

@article{jagtap2020formal,
  title={Formal synthesis of stochastic systems via control barrier certificates},
  author={Jagtap, Pushpak and Soudjani, Sadegh and Zamani, Majid},
  journal={IEEE Trans. Autom. Control},
  volume={66},
  number={7},
  pages={3097--3110},
  year={2020},
  publisher={IEEE}
}

@book{belta2017formal,
  title={Formal methods for discrete-time dynamical systems},
  author={Belta, Calin and Yordanov, Boyan and Gol, Ebru Aydin},
  volume={89},
  year={2017},
  publisher={Springer}
}

@book{tabuada2009verification,
  title={Verification and control of hybrid systems: a symbolic approach},
  author={Tabuada, Paulo},
  year={2009},
  publisher={Springer Science \& Business Media}
}

@article{lavaei2021compositional,
  title={Compositional abstraction-based synthesis of general MDPs via approximate probabilistic relations},
  author={Lavaei, Abolfazl and Soudjani, Sadegh and Zamani, Majid},
  journal={Nonlinear Anal. Hybrid Syst.},
  volume={39},
  pages={100991},
  year={2021},
  publisher={Elsevier}
}

@inproceedings{wang2025unraveling,
  author    = {Wang, Ruohan and Sun, Zhiyong and Haesaert, Sofie },
  title     = {Unraveling Tensor Structures in Correct-by-Design Controller Synthesis},
  booktitle = {64th Conf. Decis. Control},
  pages={to appear},
  year      = {2025},
  organization = {IEEE}
}

@article{liu2016finite,
  title={Finite abstractions with robustness margins for temporal logic-based control synthesis},
  author={Liu, Jun and Ozay, Necmiye},
  journal={Nonlinear Anal. Hybrid Syst.},
  volume={22},
  pages={1--15},
  year={2016},
  publisher={Elsevier}
}

@article{fainekos2009robustness,
  title={Robustness of temporal logic specifications for continuous-time signals},
  author={Fainekos, Georgios E and Pappas, George J},
  journal={Theor. Comput. Sci.},
  volume={410},
  number={42},
  pages={4262--4291},
  year={2009},
  publisher={Elsevier}
}

@article{girard2007approximation,
  title={Approximation metrics for discrete and continuous systems},
  author={Girard, Antoine and Pappas, George J},
  journal={IEEE Trans. Autom. Control},
  volume={52},
  number={5},
  pages={782--798},
  year={2007}
}

@article{desharnais2004metrics,
  title={Metrics for labelled Markov processes},
  author={Desharnais, Josee and Gupta, Vineet and Jagadeesan, Radha and Panangaden, Prakash},
  journal={Theor. Comput. Sci.},
  volume={318},
  number={3},
  pages={323--354},
  year={2004},
  publisher={Elsevier}
}

@article{cai2025safety,
  title={Safety-Critical Learning of Robot Control with Temporal Logic Specifications},
  author={Cai, Mingyu and Vasile, Cristian-Ioan},
  journal={IEEE Trans. Autom. Control},
  year={2025},
  publisher={IEEE}
}

@book{bertsekas1996stochastic,
  title={Stochastic optimal control: the discrete-time case},
  author={Bertsekas, Dimitri and Shreve, Steven E},
  volume={5},
  year={1996},
  publisher={Athena Scientific}
}

@inproceedings{van2023syscore,
  title={SySCoRe: Synthesis via stochastic coupling relations},
  author={Van Huijgevoort, Birgit and Sch{\"o}n, Oliver and Soudjani, Sadegh and Haesaert, Sofie},
  booktitle={Proceedings of the 26th ACM HSCC},
  pages={1--11},
  year={2023}
}

@misc{wang2025robusttensor,
  author       = {Ruohan Wang and Siyuan Liu and Zhiyong Sun and Sofie Haesaert},
  title        = {Correct-by-Design Control Synthesis of Stochastic Multi-agent Systems: a Robust Tensor-based Solution},
  year         = {2025},
  note         = {Preprint. Available at: \url{https://bit.ly/4osQPY8}}
}

@article{wiltz2025parallelized,
  title={Parallelized robust distributed model predictive control in the presence of coupled state constraints},
  author={Wiltz, Adrian and Chen, Fei and Dimarogonas, Dimos V},
  journal={Automatica},
  volume={171},
  pages={111952},
  year={2025},
  publisher={Elsevier}
}

@inproceedings{vlahakis2024probabilistic,
  title={Probabilistic tube-based control synthesis of stochastic multi-agent systems under signal temporal logic},
  author={Vlahakis, Eleftherios E and Lindemann, Lars and Sopasakis, Pantelis and Dimarogonas, Dimos V},
  booktitle={63rd Conf. Decis. Control},
  pages={1586--1592},
  year={2024},
  organization={IEEE}
}

@article{wang2025distributed,
  title={Distributed safe control design and probabilistic safety verification for multi-agent systems},
  author={Wang, Han and Papachristodoulou, Antonis and Margellos, Kostas},
  journal={Automatica},
  volume={179},
  pages={112393},
  year={2025},
  publisher={Elsevier}
}

@article{tkachev2017quantitative,
  title={Quantitative model-checking of controlled discrete-time Markov processes},
  author={Tkachev, Ilya and Mereacre, Alexandru and Katoen, Joost-Pieter and Abate, Alessandro},
  journal={Inform. and Comput.},
  volume={253},
  pages={1--35},
  year={2017},
  publisher={Elsevier}
}

@article{kupferman2001model,
  title={Model checking of safety properties},
  author={Kupferman, Orna and Vardi, Moshe Y},
  journal={Form. Methods Syst. Des.},
  volume={19},
  number={3},
  pages={291--314},
  year={2001},
  publisher={Springer}
}

@article{de2013linear,
  title={Linear temporal logic and linear dynamic logic on finite traces},
  author={De Giacomo, Giuseppe and Vardi, Moshe Y},
  year={2013},
  publisher={Association for Computing Machinery}
}
\newpage
\section{Appendix}
\begin{definition}
    [Coupling similar models \cite{haesaert2020robust}] \label{def:coupled_similar_model}
    \hspace{1mm} Suppose gMDPs $\mdpM_1,\mdpM_2\in\mdpM_{\spaceY}$ are given with lifted Borel measurable stochastic kernel $\mathbb{W}_{\Tr}$ for a given relation $\relation$ as defined in the beginning of Section~\ref{subsec:asr}. We define the coupling gMDP
    \begin{align}
        \mdpM_1 ||_{\relation} \mdpM_2 := (\spaceX_{||},\spaceU_{||},\spaceY_{||},x_{0||},\Tr_{||},h_{||})
    \end{align}
    with
    \begin{itemize}
        \item $\spaceX_{||}:=\spaceX_1 \times \spaceX_2$ the combined state space;
        \item $\spaceU_{||}:=\spaceU_1$ the input of system $\mdpM_1$;
        \item $\spaceY_{||}:=\spaceY$ the common output space;
        \item $x_{0||}:=(x_{1,0},x_{2,0})\in\spaceX_{||}$ the pair of initial states of $\mdpM_1$ and $\mdpM_2$;
        \item $\Tr_{||}:=\mathbb{W}_{\Tr}$ the stochastic transition kernel;
        \item $h_{||}(x_1,x_2):=h_2(x_2)$ the output mapping of $\mdpM_2$.
    \end{itemize}    
\end{definition}
\begin{proposition}[Control mapping from coupled system to concrete system \cite{haesaert2020robust}]  \hspace{1mm}\label{prop:control_refinement_back}
 Suppose $\mdpM_1,\mdpM_2\in\mdpM_{\spaceY}$ with $\mdpM_1 \preceq_0^\delta \mdpM_2$. For every control strategy $\boldsymbol{C}_1$ for $\mdpM_1 ||_{\relation} \mdpM_2$, there exists a control strategy $\boldsymbol{C}_2$ for $\mdpM_2$, such that their respective probability distribution over the space of output traces are exactly the same.
\end{proposition}
\noindent {\bfseries Proof of $d_{\mathbb Y}(\hat y,y):=\max_i (d_{\mathbb Y_i}(\hat y_i,y_i))$ \eqref{eq:max_distance_metric} is a metric.}
\begin{proof}
 {\bfseries (Non-negativity)} \hspace{1mm} Given that $d_{\spaceY_i}$ is a metric, i.e., 
\begin{align}
    d_{\spaceY_i}(\hat{y}_i,y_i) \geq 0,\quad \forall i \in \{1,\ldots,N\}, \nonumber
\end{align}
it is trivial that $\max_i (d_{\spaceY_i}(\hat{y}_i,y_i)) \geq 0$.
\par
\noindent {\bfseries (Identity)} \hspace{1mm} 
Given that $d_{\mathbb Y}(\hat y,y):=\max_i (d_{\spaceY_i}(\hat{y}_i,y_i))=0$ and the proved non-negativity of $d_{\mathbb Y}(\hat y,y)$, we have
\begin{align}
    d_{\spaceY_i}(\hat{y}_i,y_i) =0 ,\quad \forall i \in \{1,\ldots,N\}. \nonumber
\end{align}
Since $d_{\spaceY_i}$ is a metric, it follows trivially that $\hat{y}_i=y_i$ for all $i$, and that $\hat{y}=y$ where $\hat{y}=\prod_{i=1}^N \hat{y}_i$ and $y=\prod_{i=1}^{N} y_i$.
\par
Respectively, given that $\hat{y}=y$ with $\hat{y}$ defined as $\hat{y}=\prod_{i=1}^N \hat{y}_i$ and $y$ defined as $y=\prod_{i=1}^{N} y_i$, it is obvious that
\begin{align}
    \hat{y}_i=y_i,\quad \forall i \in \{1,\ldots,N\}. \nonumber
\end{align}
Since $d_{\spaceY_i}$ is a metric, i.e.,
\begin{align}
    d_{\spaceY_i}(\hat{y}_i,y_i) =0 ,\quad \forall i \in \{1,\ldots,N\}. \nonumber
\end{align}
This equality then follows naturally
\begin{align}
    \max_i (d_{\spaceY_i}(\hat{y}_i,y_i))=0. \nonumber
\end{align}

\par
\noindent {\bfseries (Symmetry)} \hspace{1mm} Since $d_{\spaceY_i}$ is a metric, i.e.,
\begin{align}
    d_{\spaceY_i}(\hat{y}_i,y_i)=d_{\spaceY_i}(y_i,\hat{y}_i), \quad \forall i \in \{1,\ldots,N\}, \nonumber
\end{align}
it is trivial that $\max_i(d_{\spaceY_i}(\hat{y}_i,y_i)) = \max_i(d_{\spaceY_i}(y_i,\hat{y}_i))$, and that $ d_{\spaceY}(\hat{y},y)= d_{\spaceY}(y,\hat{y}).$

\par
\noindent {\bfseries (Triangular inequality)}\hspace{1mm} $d_{\spaceY_i}$ is a metric, i.e.,
\begin{align} \label{eq:proof_metric_inequa}
    d_{\spaceY_i}(\hat{y}_i,y_i)\leq d_{\spaceY^i}(\hat{y}_i,z_i)+d_{\spaceY_i}(z_i,y_i), \quad \forall i \in \{1,\ldots,N\}. 
\end{align}
We maximize two items over $i$ in the right-hand side of the inequality \eqref{eq:proof_metric_inequa} and obtain following inequality holding for all $i\in \{1,\ldots,N\}$:
\begin{equation} \label{eq:proof_metric_inequa2}
\begin{aligned}
    d_{\spaceY_i}(\hat{y}_i,y_i) &\leq d_{\spaceY_i}(\hat{y}_i,z_i)+d_{\spaceY_i}(z_i,y_i) 
    \\
    &\leq \max_i(d_{\spaceY_i}(\hat{y}_i,z_i))+\max_i(d_{\spaceY_i}(z_i,y_i)).
\end{aligned}
\end{equation}
Since \eqref{eq:proof_metric_inequa2} holds true for all $i$, it is trivial that
\begin{align}
    \max_i(d_{\spaceY_i}(\hat{y}_i,y_i)) \leq \max_i(d_{\spaceY_i}(\hat{y}_i,z_i))+\max_i(d_{\spaceY_i}(z_i,y_i)).
    \nonumber
\end{align}
\end{proof}

\noindent {\bfseries Proof of Proposition~\ref{prop:fht}}
\begin{proof}
    \eqref{eq:fht_inequa} trivially holds for $q=q_f$ and $q=q_s$. Therefore, in the rest of our proof, we are only interested in $q\in Q \setminus \{q_f,q_s\}$. We will prove it by induction. Based on Proposition~\ref{prop:control_refinement_back}, we in equivalence characterize the left-hand side of \eqref{eq:fht_inequa} using value functions defined for coupled gMDP $\hat{\mdpM}||_{\relation} \mdpM$, i.e.,
    \begin{align} \label{eq:proof_satprob_on_coupled_similar_tree}
        \mbP^{x_0}_{\boldsymbol{\pi} \times \mdpM_\DFAs}(\exists t \leq l: \bar{x}_t \in \bar{\spaceX}_f ):=\tensorV^l(\hat{x}_0,x_0,q). 
    \end{align}
    We implicitly expand a tree via operators $\op(\cdot)$ applied on the combined state $(\hat{x}_i,x_i)$ of agent $i$. Based on Theorem~\ref{theorem:tree_to_value_function} and the definition of DFA-informed operators, 
\begin{equation}
     \tensorV^{l=1}(\hat{x},x,q)=\left\{\begin{array}{l}
0, \quad \text{if } \{n | \mathcal{L}_Q(n)=q\}=\emptyset\\
\bigotimes_{i=1}^{N} \op_{\subphi^{i}}^{\pi_q^{i}} (\mathrm{v}^{i})(1)(\hat{x}_{i},x_{i},q_f), \quad \text{otherwise,}
\end{array}\right. \nonumber
\end{equation}
where $\mathrm{v}^{i}(1)(\hat{x}_{i},x_{i},q_f):=\boldsymbol{1}$ is the initial vertex value of the tree.
\\
We similarly expand another tree via operators $\op(\cdot)$ applied on the state of $\hat{x}_i$ of approximated agent $i$. Similarly,
\begin{equation}
     \tensorV^{l=1}(\hat{x},q)=\left\{\begin{array}{l}
0, \quad \text{if } \{n | \mathcal{L}_Q(n)=q\}=\emptyset\\
\bigotimes_{i=1}^{N} \op_{\subphi^{i}}^{\pi_q^{i}} (\mathrm{v}^{i})(1)(\hat{x}_{i},q_f), \quad \text{otherwise,}
\end{array}\right. \nonumber
\end{equation}
where $\mathrm{v}^{i}(1)(\hat{x}_{i},q_f):=\boldsymbol{1}$ is the initial vertex value of the tree.
Based on that $\mathrm{v}^{i}(1)(\hat{x}_{i},q_f)=\mathrm{v}^{i}(1)(\hat{x}_{i},x_i,q_f)=\boldsymbol{1}$, and Condition~\textbf{L3} of $\delta$-lifted relation defined in Section~\ref{subsec:asr}, the following inequality trivially holds:
\begin{align} \label{eq:proof_coupled_geq_abs}
    \op_{\subphi^{i}}^{\pi_q^{i}} (\mathrm{v}^{i})(1)(\hat{x}_{i},x_{i},q_f) \geq \op_{\subphi^{i}}^{\pi_q^{i}} (\mathrm{v}^{i})(1)(\hat{x}_{i},q_f) -\delta^i.
\end{align}
By applying the outer product operation over all agents $i=1,\ldots,N$ of both sides of the inequality in \eqref{eq:proof_coupled_geq_abs}, it is trivial that
\begin{equation}
\begin{aligned}
    \bigotimes_{i=1}^N \op_{\subphi^{i}}^{\pi_q^{i}} (\mathrm{v}^{i})(1)(\hat{x}_{i},x_{i},q_f) & \geq  \bigotimes_{i=1}^N \biggl(\op_{\subphi^{i}}^{\pi_q^{i}}(\mathrm{v}^{i})(1)(\hat{x}_{i},q_f) -\delta^i \biggr) \nonumber
    \\
    &\geq \bigotimes_{i=1}^N \op_{\subphi^{i}}^{\pi_q^{i}}(\mathrm{v}^{i})(1)(\hat{x}_{i},q_f)  -\boldsymbol{\delta},\nonumber
\end{aligned}
\end{equation}
where $\boldsymbol{\delta}:=1-\prod_{i=1}^N(1-\delta^i)$ is the probability deviation quantified for the multi-agent system as in \eqref{eq:1-prod_delta}.
\par Note that $\mathbb{P}(H_{\hat{\spaceX} \times \{q_f\}}(\hat{x},q) \geq 1)=1$, we conclude our proof of \eqref{eq:fht_inequa} holding for $l=1$. We proceed to prove that given \eqref{eq:fht_inequa} holds for $l=k$, then it holds for $l=k+1$. We assume that \eqref{eq:fht_inequa} holds for $l=k$, i.e., $\tensorV^{k}(\hat{x},x,q') \geq \tensorV^{k}(\hat{x},q')-\boldsymbol{\delta} \sum_{h=1}^{k} \mathbb{P}(H_{\hat{\spaceS} \times \{q_f\}}(\hat{x},q') \geq h)$.
Based on Theorem~\ref{theorem:tree_to_value_function}, we represent $\tensorV^{k+1}(\hat{x},x,q)$ as the summation of $\tensorV^{k}(\hat{x},x,q)$ and $q$-labeled leafs in the expanded tree, i.e., \begin{tightdisplay}
\begin{equation}\label{eq:proof_thrid_item}
\begin{aligned}
    \tensorV^{k+1}(\hat{x},x,q)&=\tensorV^{k}(\hat{x},x,q)+\sum_{\{n| n\in \mathcal{G}^{k+1}.{\text{leaves}}\}} v(n)(\hat{x},x,q)
    \\
    &\geq \tensorV^{k}(\hat{x},q')-\boldsymbol{\delta} \sum_{h=1}^{k} \mathbb{P}(H_{\hat{\spaceS} \times \{q_f\}}(\hat{x},q') \geq h) 
    \\
    & \quad \quad \quad + \sum_{\{n| n\in \mathcal{G}^{k+1}.{\text{leaves}}\}} v(n)(\hat{x},x,q).
\end{aligned} 
\end{equation} 
\end{tightdisplay}
\noindent We unravel the computation of the third item of the right-hand side of the inequality \eqref{eq:proof_thrid_item} as (with $n^{-}$ denoting the \textbf{unique} parent of $n$)
\begin{tightdisplay}
\begin{equation}
\begin{aligned}
    &\sum_{\{n| n\in \mathcal{G}^{k+1}.{\text{leaves}}\}} v(n)(\hat{x},x,q)
    \\
    &=\sum_{\{n| n\in \mathcal{G}^{k+1}.{\text{leaves}}\}} \bigotimes_{i=1}^{N} \biggl( \op_{\subphi^{i}}^{\pi_q^{i}} (\mathrm{v}^{i})(n^{-})(\hat{x}_{i},x_{i},q') \biggr)
    \\
    &\geq  \sum_{\{n| n\in \mathcal{G}^{k+1}.{\text{leaves}}\}} \bigotimes_{i=1}^{N} \biggl( \op_{\subphi^{i}}^{\pi_q^{i}} (\mathrm{v}^{i})(n^{-})(\hat{x}_{i},q') -\delta^{i} \biggr)
    \\
    &= \sum_{\{n| n\in \mathcal{G}^{k+1}.{\text{leaves}}\}}\biggl( \bigotimes_{i=1}^{N}  \op_{\subphi^{i}}^{\pi_q^{i}} (\mathrm{v}^{i})(n^{-})(\hat{x}_{i},q') +f_n(\boldsymbol{\delta})\biggr),
\end{aligned} \label{eq:replace_third_item}
\end{equation}
\end{tightdisplay}
\noindent where $f_n(\boldsymbol{\delta})$ is negative and represents the \textbf{true} probability loss for accepted word of $n$. The length of the word is $k+1$. We rewrite \eqref{eq:proof_thrid_item} based on \eqref{eq:replace_third_item} as
\begin{tightdisplay}
\begin{equation}
    \begin{aligned}
        &\tensorV^{k+1}(\hat{x},x,q)
        \\
        &\geq \tensorV^k(\hat{x},q') -\boldsymbol{\delta} \sum_{h=1}^{k} \mathbb{P}(H_{\hat{\spaceS} \times \{q_f\}}(\hat{x},q') \geq h) 
        \\
        & \quad + \sum_{\{n| n\in \mathcal{G}^{k+1}.{\text{leaves}}\}}\biggl( \bigotimes_{i=1}^{N}  \op_{\subphi^{i}}^{\pi_q^{i}} (\mathrm{v}^{i})(n^{-})(\hat{x}_{i},q') +f_n(\boldsymbol{\delta})\biggr)
        \\
        &=\tensorV^{k+1}(\hat{x},q)-\boldsymbol{\delta} \sum_{h=1}^{k} \mathbb{P}(H_{\hat{\spaceX} \times \{q_f\}}(\hat{x},q) \geq h) \\
        & \quad \quad \quad + \sum_{\{n|n\in \mathcal{G}^{k+1}.{\text{leaves}}\}} \biggl[  f_n(\boldsymbol{\delta}) \biggr]
        \\
        &=\tensorV^{k+1}(\hat{x},q)-\boldsymbol{\delta} \sum_{h=1}^{k} \mathbb{P}(H_{\hat{\spaceX} \times \{q_f\}}(\hat{x},q) \geq h) \\
        & \quad \quad \quad - \boldsymbol{\delta} \mathbb{P}(H_{\hat{\spaceX}\times \{q_f\}}(\hat{x},q)=k+1)
        \\
        &\geq \tensorV^{k+1}(\hat{x},q)-\boldsymbol{\delta} \sum_{h=1}^{k} \mathbb{P}(H_{\hat{\spaceX} \times \{q_f\}}(\hat{x},q) \geq h) \\
        & \quad \quad \quad - \boldsymbol{\delta} \mathbb{P}(H_{\hat{\spaceX}\times \{q_f\}}(\hat{x},q)\geq k+1)
        \\
        &=\tensorV^{k+1}(\hat{x},q)-\boldsymbol{\delta} \sum_{h=1}^{k+1} \mathbb{P}(H_{\hat{\spaceX} \times \{q_f\}}(\hat{x},q) \geq h).
    \end{aligned} \nonumber
\end{equation}
\end{tightdisplay}
\noindent This proves the proposition by induction.

\end{proof}

\noindent {\bfseries Proof of Theorem~\ref{theorem:branch_length_lossProb}}
\begin{proof}
    It is trivial that
    \begin{align}
        \mathbb{P}(H_{\hat{\spaceX} \times \{q_f\}}(\hat{x},q) \geq h) = 1- \mathbb{P}(H_{\hat{\spaceX} \times \{q_f\}}(\hat{x},q) < h). \nonumber
    \end{align}
    We re-write the inequality \eqref{eq:fht_inequa} in Proposition~\ref{prop:fht} as
    \begin{align}
        \tensorV^{l}(\hat{x},x,q) \geq \tensorV^{l}(\hat{x},q)-\boldsymbol{\delta} \sum_{h=1}^{l} \biggl(1-\mathbb{P}(H_{\hat{\spaceX} \times \{q_f\}}(\hat{x},q) < h) \biggr), \nonumber
    \end{align}
    where $\mathbb{P}(H_{\hat{\spaceX} \times \{q_f\}}(\hat{x},q) < h) $ denotes the probability the controlled system product reaches the accepted state within $h-1$ control steps. Based on Theorem~\ref{theorem:tree_to_value_function}, such probability can be lower bounded by the vertices' values, i.e.,
    \begin{equation}
        \begin{aligned}
        \mathbb{P}(H_{\hat{\spaceX}\times \{q_f\}}(\hat{x},q) < h) &\geq \sum_{n\in \mathcal{L}_{Q_{h-1}}^{-1} (q)} v(n)
        \\
        & = \sum_{i=0}^{h-1}\sum_{n\in\mathcal{D}_i^q}v(n)
        \label{eq:fht_geq_node}
    \end{aligned}
    \end{equation}
    where $\mathcal{L}_{Q_{h-1}}^{-1} (q):Q\rightarrow \setnodes $ gives the vertices set in $\mathcal{G}_{h-1}$ which is labeled $q$, and $\mathcal{D}_i^q$ is the set containing $q$-labeled vertices, each of which has exactly $i$ predecessors, defined as in \eqref{eq:same_predecessor}.
\end{proof}


\begin{lemma}
\label{lemma:operator_robust_nonrobust} \hspace{1mm}
    Suppose $\hat{\mdpM}^{i} \preceq_0^{\delta^{i}} \mdpM^{i}$ with simulation relation $\relation_{i}$ as defined in Definition~\ref{def:simulation-relation} and a mapping $\pi_q^{i}: \hat{\spaceX}_{i} \rightarrow \hat{\spaceU}_{i}$ for all $i\in \{1,\ldots,N\}$ is given.
    Let the tensor components of $v(\hat{x},x,q)$ and
    $v(\hat{x},q)$ have following inequality
    \begin{align}
        \mathrm{v}(\hat{x}_{i},x_{i},q) \geq \mathrm{v}(\hat{x}_{i},q), \quad \forall (\hat{x}_i,x_i)\in\relation_i, \quad \forall i\in \{1,\ldots,N\}.
    \end{align}
    
    
    Then for $(\hat{x}_{i},x_{i})\in \relation_{i}$ and for $i\in \{1,\ldots,N\}$, 
   \begin{align}
         \op^{\pi_q^{i}}_{\subphi^{i}} (\mathrm{v})(\hat{x}_{i},x_{i},q)  \geq \op^{\pi_q^{i}}_{\delta^{i},\subphi^{i}} (\mathrm{v})(\hat{x}_{i},q). \label{eq:lemma_comp_inequality}
    \end{align}

 
\end{lemma}
\par
\noindent {\bfseries Proof of Lemma~\ref{lemma:operator_robust_nonrobust}}
\begin{proof}
We begin the proof computing $\op^{\pi_q^{i}}_{\subphi^{i}} (\mathrm{v})(\hat{x}_{i},q)$:
\begin{align}  \op^{\pi_q^{i}}_{\subphi^{i}} (\mathrm{v})(\hat{x}_{i},q) := \expectation^{\hat{x}_{i}'}[\mathcal{L}_{\subphi^{i}}(\hat{x}_{i}')\mathrm{v}(\hat{x}_{i}',q) \mid \hat{x}_{i},\hat{u}_{i}=\pi_q^{i}(\hat{x}_{i})]. \label{eq:in_proof_inetr_expec}
\end{align}
For $(\hat{x}_{i},x_i)\in \relation_{i}$ and policy $\pi_q^{i}: \hat{\spaceX}_{i} \rightarrow \hat{\spaceU}_{i}, \forall i\in\{1,\ldots,N\}$ applied to the lifted kernel of the composed system $\hat{\mdpM}^{i} ||_{\relation_{i}} \mdpM^{i}$, condition \textbf{L1} \eqref{eq:delta_lift_L1} gives the equivalent integral to the expectation in \eqref{eq:in_proof_inetr_expec} as
\begin{equation}
    \begin{aligned}
        &
        \int_{\hat{\spaceX}_{i} \times \spaceX_{i}}\mathcal{L}_{\subphi^{i}}(\hat{x}_{i}')\mathrm{v}(\hat{x}_{i}',q) \mathbb{W}_{\Tr^{i}}(d\hat{x}_{i}' \times dx_{i}' | \pi_q^{i},\hat{x}_{i},x_i
        ) 
        \\
        =&\int_{ \relation_{i}}\mathcal{L}_{\subphi^{i}}(\hat{x}_{i}')\mathrm{v}(\hat{x}_{i}',q) \mathbb{W}_{\Tr^{i}}(d\hat{x}_{i}' \times dx_{i}' | \pi_q^{i},\hat{x}_{i},x_i
        )  
        \\
        &\quad + \int_{(\hat{\spaceX}_{i} \times \spaceX_{i}) \setminus \relation_{i}}\mathcal{L}_{\subphi^{i}}(\hat{x}_{i}')\mathrm{v}(\hat{x}_{i}',q) \mathbb{W}_{\Tr^{i}}(d\hat{x}_{i}' \times dx_{i}' | \pi_q^{i},\hat{x}_{i},x_i
        ) 
        \\
        \leq & \int_{ \relation_{i}}\mathcal{L}_{\subphi^{i}}(\hat{x}_{i}')\mathrm{v}(\hat{x}_{i}',q) \mathbb{W}_{\Tr^{i}}(d\hat{x}_{i}' \times dx_{i}' | \hat{u}_{i},\hat{x}_{i},x_i
        )  + \delta^{i}.
    \end{aligned} \nonumber
\end{equation}
The inequality holds due to $\mathbb{W}_{\Tr^{i}}\biggl(  (\hat{\spaceX}_{i} \times \spaceX_{i}) \setminus \relation_{i} | \hat{u}_{i},\hat{x}_{i},x_i\biggr) \leq \delta^{i}$. According to the assumption of the lemma, i.e., $\mathrm{v}(\hat{x}_{i},x_i,q) \geq \mathrm{v}(\hat{x}_{i},q)$ for all $(\hat{x}_{i},x_i)\in \relation_{i}$, and for all $i$, the integral over $\relation_{i}$ is equal or smaller to 
\begin{equation}
    \begin{aligned}
        &\int_{ \relation_{i}}\mathcal{L}_{\subphi^{i}}(\hat{x}_{i}')\mathrm{v}(\hat{x}_{i}',x_{i}',q) \mathbb{W}_{\Tr^{i}}(d\hat{x}_{i}' \times dx_{i}' | \hat{u}_{i},\hat{x}_{i},x_i
        )
        \\
        \leq & \int_{\hat{\spaceX}_{i}\times \spaceX_{i}}\mathcal{L}_{\subphi^{i}}(\hat{x}_{i}')\mathrm{v}(\hat{x}_{i}',x_{i}',q) \mathbb{W}_{\Tr^{i}}(d\hat{x}_{i}' \times dx_{i}' | \hat{u}_{i},\hat{x}_{i},x_i)
        \\
        =& \op^{\pi_q^{i}}_{\subphi^{i}} (\mathrm{v})(\hat{x}_{i},x_i,q).
    \end{aligned} \nonumber
\end{equation}
We have proved $\op^{\pi_q^{i}}_{\subphi^{i}} (\mathrm{v})(\hat{x}_{i},x_i,q) \geq \op^{\pi_q^{i}}_{\subphi^{i}} (\mathrm{v})(\hat{x}_{i},q)-\delta^{i}$, based on which we have
\begin{align}
            \trunc \biggl( \op^{\pi_q^{i}}_{\subphi^{i}} (\mathrm{v})(\hat{x}_{i},x_i,q) \biggr) \geq \trunc \biggl( \op^{\pi_q^{i}}_{\subphi^{i}} (\mathrm{v})(\hat{x}_{i},q)-\delta^{i} \biggr), \nonumber
\end{align}
and then, $\forall (\hat{x}_{i},x_i) \in \relation_{i},\quad \forall i\in \{1,\ldots,N\}$
\begin{align}
\op^{\pi_q^{i}}_{\subphi^{i}} (\mathrm{v})(\hat{x}_{i},x_i,q)   \geq \op^{\pi_q^{i}}_{\delta^{i},\subphi^{i}} (\mathrm{v})(\hat{x}_{i},q)  \nonumber
\end{align}
since $\op^{\pi_q^{i}}_{\subphi^{i}} (\mathrm{v})(\hat{x}_{i},x_i,q) $ trivially takes value in $[0,1]$. 


\end{proof}

\noindent {\bfseries Proof of Theorem~\ref{theorem:satprob_lb_robusttree}}
\begin{proof}
    We characterize the left-hand side of \eqref{eq:theorem:satprob_lb_robusttree} as the value functions computed based on the implicitly expanded tree $\mathcal{G}$ for coupled gMDP $\hat{\mdpM}||_{\relation} \mdpM$ based on Proposition~\ref{prop:control_refinement_back}, i.e.,
    \begin{equation}
    \begin{aligned}
        \mbP^{x_0}_{\boldsymbol{\pi} \times \mdpM_\DFAs}(\exists t \leq l: \bar{x}_t \in \bar{\spaceX}_f )&:=\tensorV^l(\hat{x}_0,x_0,q) 
        \\
        &:=\sum_{n\in \mathcal{L}_{Q_l}^{-1}(\bar{q}_0)} v_l(n),
    \end{aligned}
    \end{equation}
    with $\bar{q}_0:=\tau_\DFA(q_0,L(g(\hat{x}_0)))$ and $n$ the vertices of tree $\mathcal{G}_{\hat{\mdpM} ||_\relation \mdpM }$.
    The right-hand side of \eqref{eq:theorem:satprob_lb_robusttree} are the summation of nodes of the expanded robust tree $\mathcal{G}^{\delta}$ for approximation $\hat{\mdpM}$. Since $\mathcal{G}$ and $\mathcal{G}^{\delta}$ are initialized as 
    \begin{align}
        \mathrm{v}^i(1)(\hat{x}_i,x_i,q_f) = 1, \nonumber
    \end{align}
    and 
    \begin{align}
        \mathrm{v}^i(1)(\hat{x}_i,q_f)=1. \nonumber
    \end{align}
    Based on Lemma~\ref{lemma:operator_robust_nonrobust}, this follows naturally:
    \begin{align}
        \op_{\subphi^i}^{\pi_q^i}(\mathrm{v}^i)(1)(\hat{x}_i,x_i,q_f) \geq \op_{\delta^i,\subphi^i}^{\pi_q^i}(\mathrm{v}^i)(1)(\hat{x}_i,q_f). \nonumber
    \end{align}
    It is then trivial that
    \begin{equation}
    \begin{aligned}
        \bigotimes_{i=1}^{N}\op_{\subphi^i}^{\pi_q^i}(\mathrm{v}^i)(n)(\hat{x}_i,x_i,q) &\geq \bigotimes_{i=1}^{N} \op_{\delta^i,\subphi^i}^{\pi_q^i}(\mathrm{v}^i)(n)(\hat{x}_i,q) ,\quad \forall q,n;
        \vspace{2mm}
        \\
        v(q,n) &\geq v^\delta(q,n), \quad \forall q,n;
        \vspace{2mm}
        \\
        \sum_{n\in \mathcal{L}_{Q_l}^{-1}(q)} v_l(n) &\geq \sum_{n\in {\qmapping}_l^{-1}(q)}v_l^{\delta}(n), \quad \forall q.
    \end{aligned}\nonumber
    \end{equation}

\end{proof}

\end{document}